\documentclass{rstransa}

\usepackage{color}
\newcommand{\red}[1]{{\color{red} #1}}  

\jname{rsta}
\Journal{Phil. Trans. R. Soc}
\begin{document}

\title{Prospective Sensitivities of Atom Interferometers to Gravitational Waves and Ultralight Dark Matter}
\author{Leonardo~Badurina~$^1$, Oliver~Buchmueller~$^2$, John~Ellis~$^{1,3}$,
Marek~Lewicki~$^4$, Christopher~McCabe~$^1$, Ville~Vaskonen~$^5$}
\address{\small
$^1$~Physics Department, King's College London, Strand, London WC2R~2LS, UK\\
$^2$~High Energy Physics Group, Blackett Laboratory, Imperial College, Prince Consort Road, London SW7~2AZ, UK\\
$^3$~Theoretical Physics Department, CERN, 1211 Geneva 23, Switzerland;\\
National Institute of Chemical Physics and Biophysics, R{\" a}vala 10, 10143 Tallinn, Estonia\\
$^4$~Faculty of Physics, University of Warsaw, \\
ul. Pasteura 5, 02-093 Warsaw, Poland\\
$^5$~Institut de Fisica d’Altes Energies (IFAE), \\
Barcelona Institute of Science and Technology, Campus UAB, 08193 Bellaterra (Barcelona), Spain}

\subject{Astrophysics, cosmology, relativity, particle physics}
\corres{John Ellis\\
\email{John.Ellis@cern.ch}}

\begin{abstract}
\small{We survey the prospective sensitivities of
terrestrial and space-borne atom interferometers (AIs) to
gravitat- ional waves (GWs) generated by cosmological
and astrophysical sources, and to ultralight dark matter. We discuss the
backgrounds from gravitational gradient noise (GGN) in terrestrial detectors,
and also binary pulsar and asteroid backgrounds in space-borne detectors.
We compare the sensitivities of LIGO and LISA with those of the 
100m and 1km stages of the AION 
terrestrial AI project, as well as two options for the proposed AEDGE 
AI space mission with cold atom clouds either inside or outside the 
spacecraft, considering as possible sources the mergers of black holes and
neutron stars, supernovae, phase transitions in the early Universe, cosmic strings and quantum fluctuations in the early Universe that could have generated primordial black holes. We also review the capabilities of AION and AEDGE for detecting coherent waves of ultralight scalar dark matter.\\
~~\\
AION-REPORT/2021-04\\
KCL-PH-TH/2021-61, CERN-TH-2021-116}
\end{abstract}

\maketitle

\section{Introduction}
\label{sec:intro}

The terrestrial laser interferometer experiments LIGO and Virgo have initiated~\cite{LIGOScientific:2016aoc} a rich programme
of direct measurements of gravitational waves (GWs) with frequencies $\sim 10 - 10^3$~Hz produced by
astrophysical sources including binary systems of black holes and neutron stars. The LIGO/Virgo
observations have already raised many interesting questions in fundamental physics
as well as cosmology and astrophysics. For example, are the LIGO/Virgo black holes of astrophysical or primordial
origin? Are there black holes in the mass gap caused by pair instabilities in massive
stars? If so, do they originate from previous mergers, or some primordial mechanism, or are
they made possible by some new physics beyond the Standard Model? These are among the many
questions to be studied in future data from LIGO and Virgo, which have now been joined
by KAGRA~\cite{KAGRA:2020cvd}, with LIGO-India~\cite{Saleem:2021iwi} joining in the future. Other laser interferometer projects operating
in a similar frequency range, such as the Einstein Telescope (ET)~\cite{Maggiore:2019uih} and Cosmic explorer (CE)~\cite{Reitze:2019iox},
are also being planned.

The planned future space-borne laser interferometer experiments LISA~\cite{LISA:2017pwj},
TianQin~\cite{TianQin:2020hid} and Taiji~\cite{Taiji} will be most sensitive to GWs in a lower 
range of frequencies $\sim 10^{-3} - 10^{-2}$~Hz, where one of the primary targets will be mergers involving the
supermassive black holes (SMBHs) that infest galactic centres. Their existence begs the question how
they formed, e.g., by hierarchical mergers of intermediate-mass black holes (IMBHs) beyond the pair-instability
mass gap?

Pulsar timing arrays (PTAs) are most sensitive to GWs at much lower
frequencies $\sim 10^{-8}$~Hz. NANOGrav has recently reported evidence for a stochastic common-spectrum
process that might be due to GWs~\cite{NANOGrav:2020bcs}. If so, what might be its origin? Could it be due to SMBH binaries, the
default astrophysical interpretation, or might it be due to cosmic strings or primordial black holes?
In the short term, the NANOGrav results will be followed up by other PTAs,
and in the longer term by SKA~\cite{Combes:2021xez}.

As is apparent from this brief survey, there are gaps in the coverage of the GW frequency spectrum that
are not covered by operating or planned experiments. These include a mid-frequency gap in the band
$\sim 10^{-2} - 10$~Hz and a low-frequency gap in the range $\sim 10^{-7} - 10^{-4}$~Hz, to which might
be added the high-frequency band $\gtrsim 10^3$~Hz. Experience with the electromagnetic spectrum shows
the desirability of full-spectrum coverage, both to uncover new scientific opportunities and for synergies
with detectors in different frequency ranges. This consideration has motivated many proposals for experiments
optimized for sensitivity in the mid-frequency band $\sim 10^{-2} - 10$~Hz, such as DECIGO~\cite{Kawamura:2020pcg}
and BBO~\cite{Corbin:2005ny}.

Several atom interferometer (AI) projects target this mid-frequency band, including MAGIS~\cite{Abe:2021ksx}, MIGA~\cite{Canuel:2017rrp}, ELGAR~\cite{Canuel:2019abg}
and ZAIGA~\cite{Zhan:2019quq} as well as the terrestrial AION~\cite{Badurina:2019hst} and space-borne AEDGE~\cite{AEDGE:2019nxb} projects that we focus on in this paper.
The MAGIS Collaboration is currently constructing a 100m detector at Fermilab, and proposes to construct
a 1km detector in the Sanford Underground Research Facility (SURF).
The AION project has several stages, starting with AION-10, a 10m detector to be constructed in Oxford that has 
received initial funding, to be followed by AION-100, a 100m detector for which sites in the UK and at CERN are
under consideration, and finally a km-scale detector, AION-km. These would be followed by the AEDGE experiment, which
would consist of two satellites in medium earth orbit, nominally 40000~km apart. The AEDGE concept was
presented to the ESA as part of its Voyage 2050 forward look, and ESA is now developing a plan of
action to develop cold atom technology for deployment in space. 

The first 10m of AION is aimed at ultralight dark
matter, and is not expected to detect GWs. Our attention here is focused on the GW capabilities of
AION-100, AION-km and AEDGE, which have been discussed previously in~\cite{Badurina:2019hst} and~\cite{AEDGE:2019nxb}, respectively. This
article updates those discussions, including several new technical and scientific developments. 
We discuss the prospective GW sensitivities of these AIs in resonance and broadband mode. In the case of AION,
we review in more detail than previously the expected gravity gradient noise (GGN).
In the case of AEDGE, we discuss the astrophysical backgrounds from unresolved galactic and extragalactic
binaries, as well as from asteroids at lower frequencies below the sensitive range of LISA.
Moreover, whereas the previous studies used the optimal response function~\cite{Graham:2016plp}, for the results presented here we take into account the full response including an all-sky average of the antenna pattern of the device. Including these improvements enables us to extend the range of frequencies of our analysis compared to previous predictions, as the cuts imposed previously were designed to mimic the possible effects of noise sources.
In the previous discussion of the GW capabilities of AEDGE, we assumed that the atom clouds would be
contained within the satellites. However, it has also been suggested that one might be able to operate an AI
with clouds outside the satellites~\cite{Dimopoulos:2008sv}, and we also consider here the scientific advantages of such an option, called AEDGE+, 
without entering into a detailed technical discussion of its technical feasibility.

The GW science topics we develop in this paper include a reconsideration of the mergers of IMBHs, to which we
add a discussion of extreme mass-ratio infall (EMRI) events involving IMBHs and neutron stars. We pay
particular attention to the extended range of masses that would be opened up by possible measurements in
the low-frequency band below LISA's range of sensitivity. We also consider the possibility of
measuring GWs from a supernova, due to the gravitational memory effect generated by anisotropic
neutrino emission. In addition to these astrophysical topics, we revisit the sensitivities of AION
and AEDGE to the stochastic GW background (SGWB) from a first-order cosmological phase transition, presenting
the model-independent sensitivities of AION and AEDGE in $(T_*, \alpha)$ planes (where $T_*$ is the reheating temperature and $\alpha$ is the strength of the
transition) for various durations of the transition. We also revisit the SGWB that would be generated by cosmic strings, motivated by the
explanation they could provide~\cite{Ellis:2020ena} for the possible NANOGrav GW signal and discussing the insights into the
expansion history of the early Universe that could be provided by measurements of this SGWB. We also discuss the AION and AEDGE sensitivities to the GWs that would be generated by quantum fluctuations in the early Universe capable of generating primordial black holes.

Another interesting topic for future AIs is the
search for ultralight dark matter, and we also review the capabilities of the various stages of AION and AEDGE for detecting coherent waves of ultralight scalar dark matter, which extend many orders of magnitude beyond
present detectors.

\section{Landscape of Atom Interferometer Experiments}
\label{sec:expts}

Atom Interferometry (AI) is an established quantum sensor concept
based on the superposition and interference of atomic wave packets.
AI experimental designs take advantage of features used by state-of-the-art atomic clocks in combination with established techniques for building inertial sensors. 
For an introduction to the general concept of atom interferometry and some applications, see~\cite{Overstreet, PhysRevLett.125.191101, cite-key, Parker191, Fixler74}. 

The experimental landscape of AI projects has expanded significantly in recent years, with several terrestrial AIs based on different Cold Atom technologies currently under construction, planned or proposed. 

Four large-scale prototype projects are funded and currently under construction, namely the AION-10 in the UK, MAGIS-100 in the US, MIGA in France, and ZAIGA in China. These will demonstrate the feasibility of AI at macroscopic scales, paving the way for terrestrial km-scale experiments as the next steps. There are projects to build one or several km-scale detectors, including AION-km at the STFC Boulby facility in the UK,  MAGIA-advanced and ELGAR in Europe, MAGIS-km at the Sanford Underground Research facility (SURF) in the US, and advanced ZAIGA in China.  It is foreseen that by about 2035 one or more km-scale detectors will have entered operation. These km-scale experiments would not only be able to explore systematically for the first time the  mid-frequency band of gravitational waves, but would also serve as the ultimate technology readiness demonstrators for a space-based mission like AEDGE that would reach the ultimate sensitivity to explore the fundamental physics goals outlined in this article.
     
Meanwhile, space-borne cold atom experiments such as CACES~\cite{Liu:2018kn}, MAIUS~\cite{MAIUS} and CAL~\cite{CAL}, as well as pathfinders for the key underlying optical technologies such as FOKUS~\cite{Lezius}, KALEXUS~\cite{Dinkelaker} and JOKARUS~\cite{PhysRevApplied.11.054068}, have already demonstrated reliable operation, and much more space experience will be gained in the coming years. In parallel, several other cold atom projects are aiming to demonstrate the general space-readiness of cold-atom technology including the scaling of the basic parameters that are required for large-scale space-borne mission concepts, e.g. the medium-scale mission concept of STE-QUEST~\cite{Tino2019}.

In summary, the perspectives for large-scale atom interferometer projects are very diverse today, with a main focus on establishing the readiness of cold atom technology for use in AI experiments to explore fundamental science. 
In the following we focus on the large-scale terrestrial project AION and the space-based AEDGE mission concept to outline the enormous science potential of AI projects.        

\subsection{A Terrestrial Project: AION}

The AION project proposes a staged approach, starting with a 10-m detector - which has received initial funding -
to be followed by 100-m and km-scale detectors~\cite{Badurina:2019hst}. The 100-m detector would be
similar to the MAGIS-100 detector under construction at Fermilab~\cite{Abe:2021ksx}, and the MAGIS Collaboration also
has in mind a follow-on km-scale detector.

\begin{figure}
\centering 
\includegraphics[width=6.5cm]{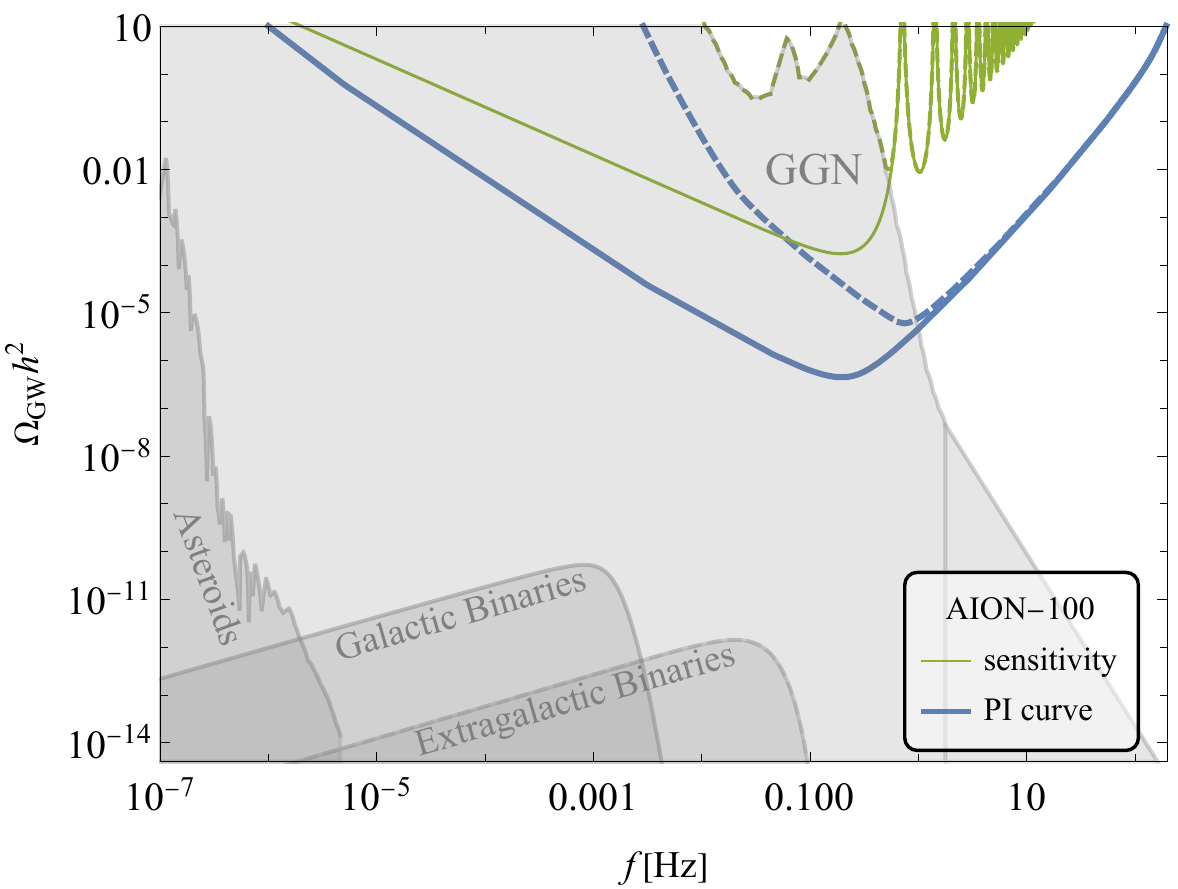}
\includegraphics[width=6.5cm]{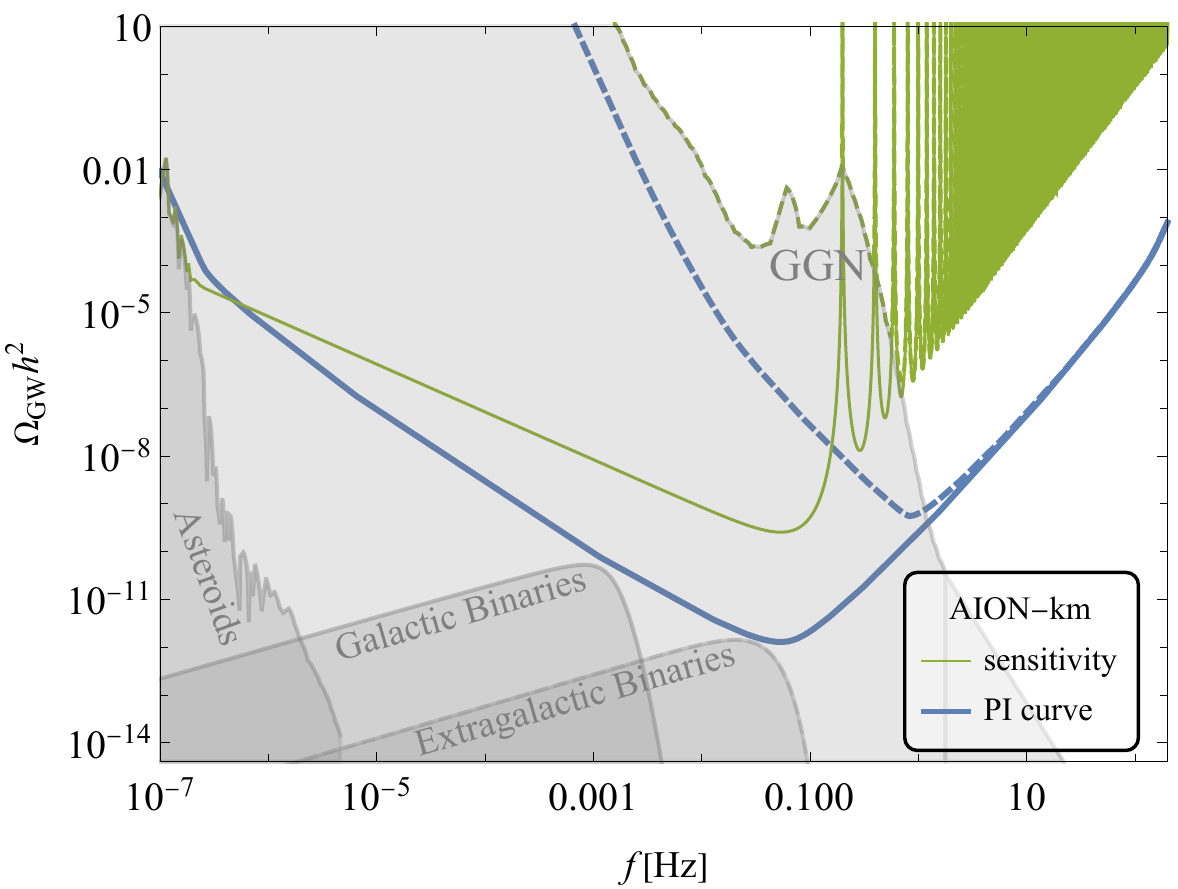}
\vspace{-3mm}
\caption{\it Sensitivities of AION-100 (left panel), 
and -km (right panel) to the density of monochromatic gravitational
waves, $\Omega_{\rm GW} h^2$. Solid green lines assume GGN can be fully mitigated while the impact of terrestrial Gravity Gradient
Noise (GGN) is indicated by dashed lines. The thick blue lines show the sensitivities to stochastic backgrounds with
power-law-integrated (PI) curves while the impact of unmitigated GGN is again shown as a dashed curve.
We also show the GGN generated by asteroids~\cite{Fedderke:2020yfy} at low frequencies, as well as the noise due to unresolved
galactic and extragalactic binaries.
}
\label{fig:AIONGGN}
\end{figure}

The AION atom interferometers are sensitive indirectly to microseismic
activity through the Newtonian gravitational effects of ground motion
on the atomic clouds, called gravity gradient noise (GGN). Measurements have been made of
the seismic noise in the NUMI shaft at Fermilab where MAGIS-100 will
be located~\cite{Abe:2021ksx}, and similar measurements have been made for the MIGA experiment~\cite{MIGAconsortium:2019efk}. The AION Collaboration plans campaigns to measure
the seismic noise at possible sites for AION-100 in the
Boulby mine and at the Daresbury Laboratory in the UK. In addition,
preliminary measurements have been made at CERN of the seismic noise on the surface and in the LHC
tunnel~\cite{CERN}, which may be indicative of the prospective levels at the top and foot 
of the PX46 shaft that is also a possible site for
AION-100.
The available measurements are in broad agreement with the New Low-Noise Model (NLNM)~\cite{GGNmodeling} 
based on measurements by the U.S. Geological Survey aimed at describing realistic quiet sites. 
In the following we show for illustration the NLNM model for GGN
also for AION-km, though there are no measurements at that depth
and the assumption of domination by Rayleigh waves would need to be validated.
Pending further investigations,
we assume that the GGN in both AION-100 and -km can be mitigated entirely by using seismometer and
other measurements. In this case, we obtain the AION-100 and -km sensitivities shown with solid lines in the left and right panels of Fig.~\ref{fig:AIONGGN}, respectively, where we also show the
backgrounds from unresolved galactic and extragalactic binaries, which do not impact the estimated AION sensitivities.

To approximate the GGN in AION we use the NLNM ~\cite{GGNmodeling} and assume simple scaling with depth 
$L$~\cite{Adamson:2018mbw,MIGAconsortium:2019efk} and noise mitigation from using $N$ interferometers~\cite{Chaibi:2016dze}: 
$S_{h {\rm GGN}}\propto 1/( \sqrt{N} L )$.
We display the resulting GGN estimates for AION-100 and -km in
the left and right panels of Fig.~\ref{fig:AIONGGN}, respectively.
We also show there the sensitivity of AION-100 and -km to monochromatic signals in terms of $\Omega_{\rm GW} h^2$ 
when operated in the fundamental mode (solid green line) and the
impact of these GGN estimates on the sensitivities 
 (dashed green lines). The experimental noise curve (solid green) is obtained as in Ref.~\cite{Badurina:2019hst}, with the improvement of the sky-averaged response function. We also show the sensitivities to a stochastic background through power-law-integrated (PI) 
 curves~\cite{Thrane:2013oya}  as thicker solid blue lines, again indicating the impact of GGN
 on these predictions by dashed blue lines. In the case of AION-km we see that
the PI sensitivity is impacted at low frequencies $\lesssim 10^{-6}$ by the
estimated GGN from asteroids~\cite{Fedderke:2020yfy}, but we emphasise that
there may also be important sources of instrumental noise at
low frequencies, as well as other sources of GGN.

\subsection{A Proposed Space-Borne Experiment: AEDGE}

We consider two possible configurations for AEDGE, based on a pair of spacecraft in medium earth orbit
with a separation of 40,000~km. One is the baseline configuration considered in~\cite{AEDGE:2019nxb}, in which
the atom clouds are contained within the spacecraft and have a size $\sim 1$~m. The other configuration
assumes atom clouds with sizes $\sim 100$~m~\cite{Dimopoulos:2008sv} that are outside the spacecraft (AEDGE+).

\begin{figure}
\centering 
\includegraphics[width=6.5cm]{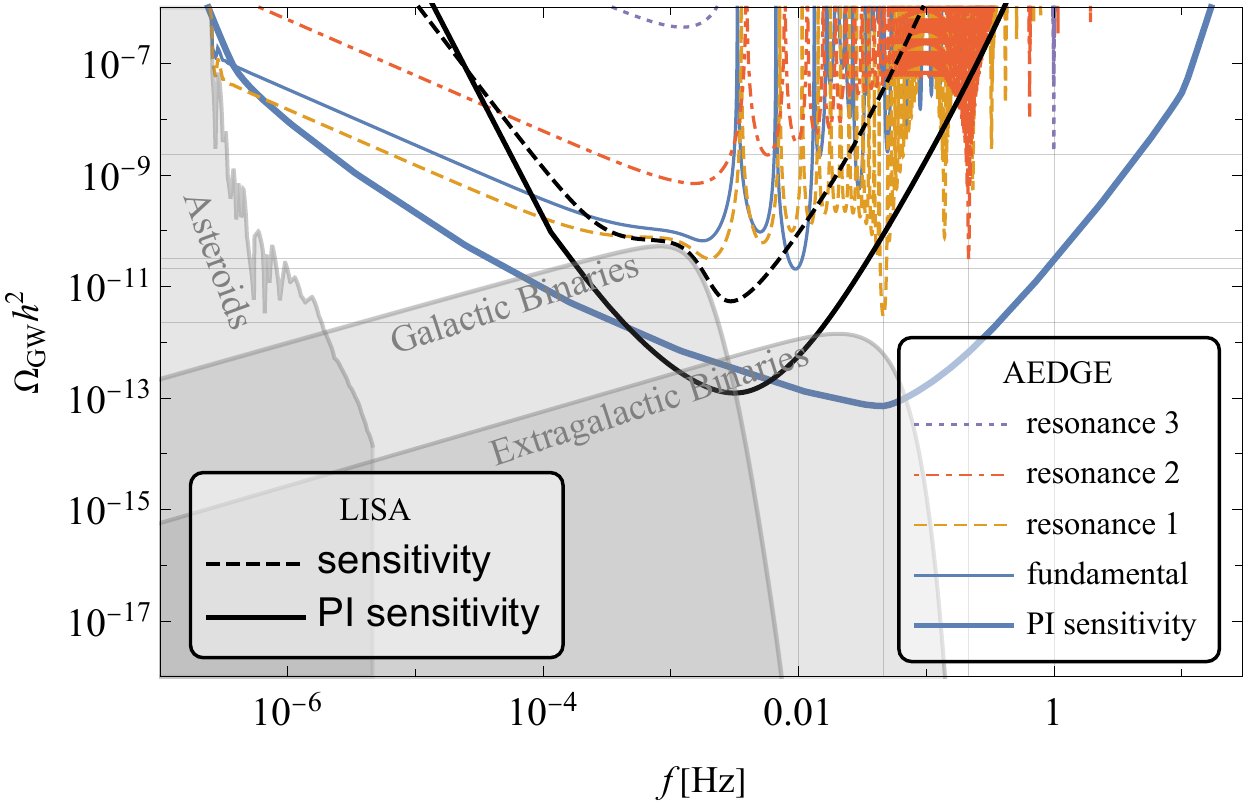}
\includegraphics[width=6.5cm]{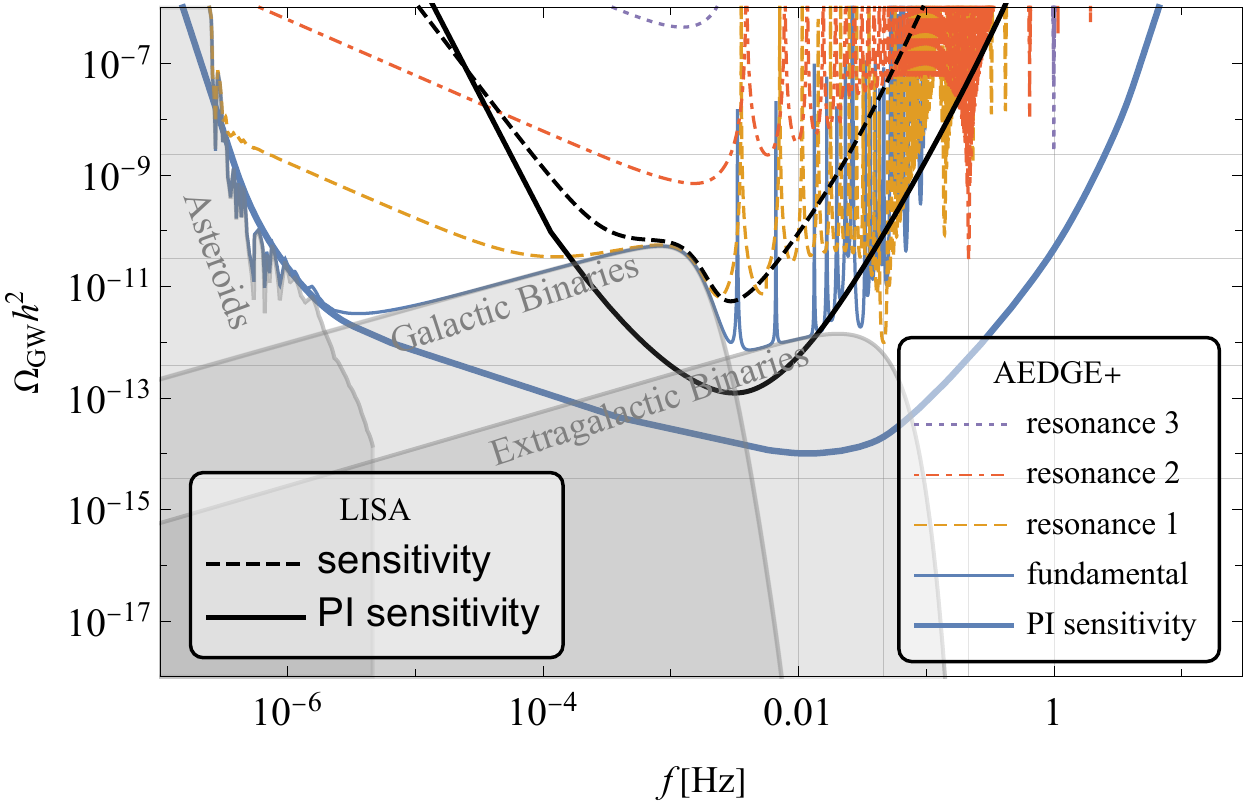}
\caption{\it As in Fig.~\ref{fig:AIONGGN}, but for the AEDGE baseline
design (left panel) and for AEDGE+ with an external 100m cloud (right panel).
Also shown are the sensitivities of LISA to monochromatic and
PI spectra.
}
\label{fig:AEDGEnoise}
\end{figure}

Analogously to the previous figures for AION, we display in Fig.~\ref{fig:AEDGEnoise} the 
$\Omega_{\rm GW} h^2$ sensitivities for the standard AEDGE configuration (left panel) and the AEDGE+ configuration with an external
cloud (right panel). We show the monochromatic sensitivities  for AEDGE operated in
the fundamental mode (solid grey lines) and in the first three resonant modes (coloured dashed and dotted
lines), including the impacts of the backgrounds due to unresolved galactic and
extragalactic binaries, as well as asteroids. 
Here, we again use the experimental noise obtained previously~\cite{AEDGE:2019nxb}, with the improvement by inclusion of the the sky-averaged response function. We also show the PI sensitivities of the two AEDGE
configurations as thicker blue lines. For comparison, we also display in both panels the monochromatic and IP sensitivities
of LISA (black dashed and solid lines, respectively). We see that the AEDGE sensitivities exceed that of
LISA in the mid-frequency band $\gtrsim 10^{-2}$~Hz, as expected, but also at lower frequencies
down to $\sim 10^{-7}$~Hz. However, the AEDGE sensitivity estimates presented here do not take into
account many possible sources of instrumental noise that were considered, e.g., in~\cite{Dimopoulos:2008sv,Hogan:2011tsw}.
We defer considerations of these noise sources for a possible future study, noting simply that the
sensitivity curves in Fig.~\ref{fig:AEDGEnoise} indicate that a space-borne AI experiment might have
interesting potential in the low-frequency band between LISA and PTA experiments, as well as
in the mid-frequency band between LISA and LIGO/Virgo/KAGRA and~ET.

\subsection{Summary of Prospective GW Sensitivities of Atom Interferometers}
\label{sec:Summary}

\begin{figure}
\centering 
\includegraphics[width=0.7\textwidth]{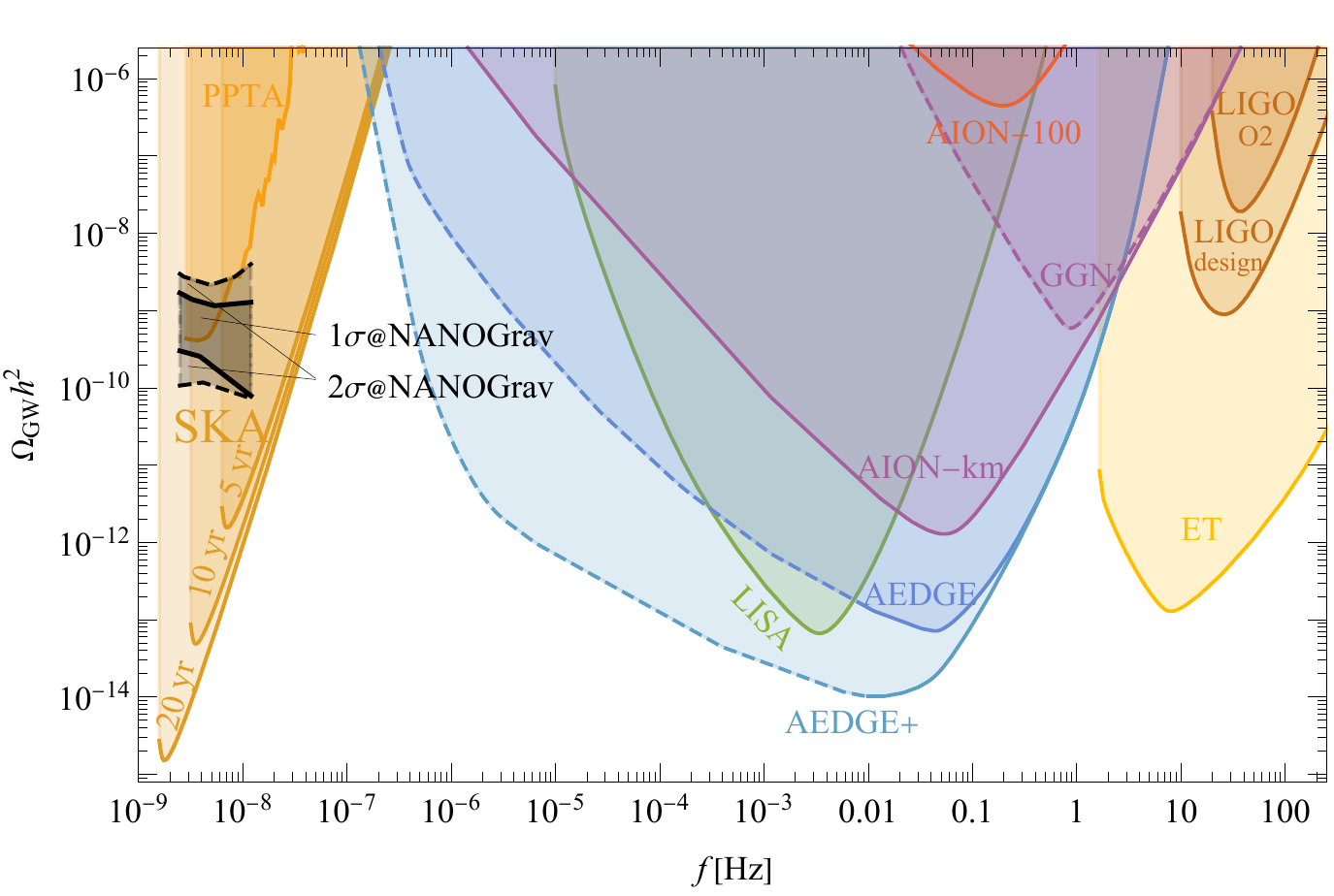}
\vspace{-3mm}
\caption{\it Comparison of the $\Omega_{\rm GW} h^2$ sensitivities to PI spectra
of AION-100, AION-km, AEDGE and AEDGE+, LIGO, ET, PTAs and SKA. Also shown are power-law fits to
the NANOGrav hint of a possible GW signal at frequencies $\lesssim 10^{-8}$~Hz. 
}
\label{fig:summary}
\end{figure}

Fig.~\ref{fig:summary} compares the possible $\Omega_{\rm GW} h^2$ sensitivities of the
AI experiments that we consider with those of other operating,
planned and proposed experiments. At low frequencies $\lesssim 10^{-7}$~Hz
we see the sensitivities of PTAs and SKA~\cite{Combes:2021xez}, at intermediate frequencies 
$\sim 10^{-2}$~Hz we see the expected LISA sensitivity~\cite{LISA:2017pwj}, 
and at higher frequencies $\gtrsim 10$~Hz
we see the sensitivity LIGO achieved during the O2 observational period and
its design goal, as well as the prospective sensitivity of the ET experiment~\cite{Maggiore:2019uih}.
These can be compared with the prospective $\Omega_{\rm GW} h^2$ sensitivities
of AION-100, AION-km, AEDGE and AEDGE+ to power-law integrated GW spectra.
For information, we also display the results from power-law fits to the NANOGrav hint~\cite{NANOGrav:2020bcs} of a possible GW signal 
at frequencies $\lesssim 10^{-8}$~Hz.

In the following Sections we touch upon some of the topics in astrophysics
and fundamental physics that could be addressed by AION and AEDGE.

\section{Astrophysical Sources of GWs}

\subsection{Black Hole/Black Hole and Black Hole/Neutron Star Mergers}
\label{sec:binaries}

The sensitivities of AION and AEDGE to binary black hole/black hole mergers were discussed
in~\cite{Badurina:2019hst} and~\cite{AEDGE:2019nxb}, respectively. Here we review and update the previous estimates for
AION commenting on the recent measurements of the seismic background at the Fermilab site of MAGIS-100
and at CERN, and update the estimates for AEDGE taking into account the backgrounds expected
from unresolved galactic and extragalactic binaries and considering the external cloud configuration.
Also, we consider asymmetric black hole/black hole mergers and black hole/neutron star mergers, 
which were not discussed in~\cite{AION,AEDGE:2019nxb}.

\begin{figure}
\centering 
\includegraphics[width=9cm]{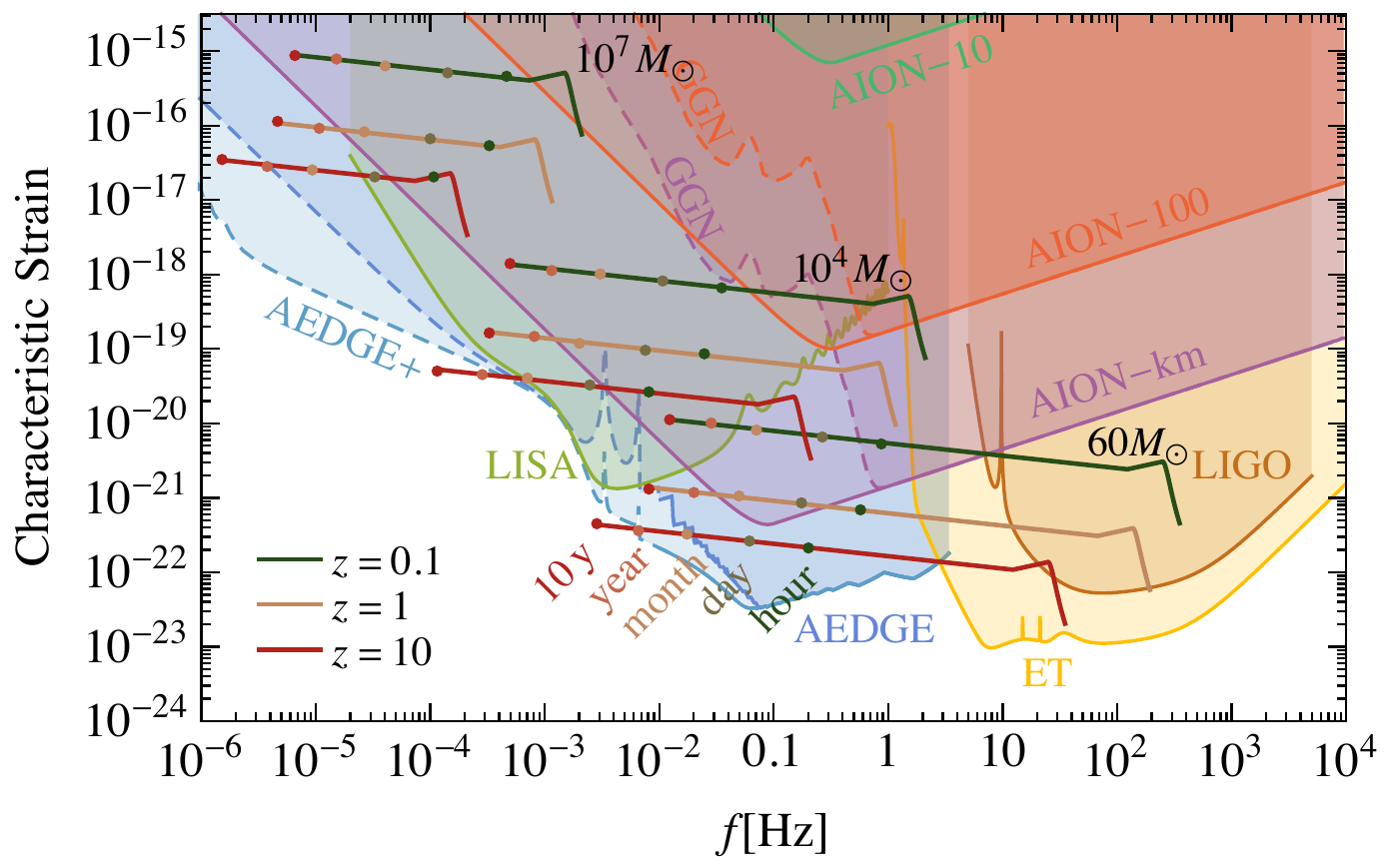}
\caption{\it Strain sensitivities of AION-10, -100 and -km, AEDGE and AEDGE+, compared
with those of LIGO, LISA and ET and the signals expected from mergers of equal-mass
binaries whose masses are $60, 10^4$ and $10^7$ solar masses.  The assumed redshifts are
$z = 0.1, 1$ and $10$, as indicated. Also shown are the remaining times during inspiral before the
final mergers.}
\label{fig:strain}
\end{figure}

Fig.~\ref{fig:strain} displays the strain sensitivities of AION-100 and -km, AEDGE and AEDGE+, 
compared with the signals expected from mergers of equal-mass
binaries with combined masses of $60, 10^4$ and $10^7$ solar masses occurring at the redshifts
$z = 0.1, 1$ and $10$, as indicated. The dashed lines correspond to the GGN level expected in the NLNM that is
consistent with seismic measurements
made at Fermilab~\cite{Abe:2021ksx} and CERN~\cite{CERN}. Strategies to mitigate these noise levels are
under investigation, which will be increasingly challenging
at lower frequencies. The solid sensitivity curves for
AION-100 and -km assume that the GGN can be completely
mitigated. The lower AEDGE sensitivity curve is for the external cloud configuration, AEDGE+, and
shows the impacts of the extragalactic and galactic binary backgrounds at frequencies ${\cal O}(10^{-2})$
and ${\cal O}(10^{-3})$, respectively.~\footnote{We stress again that the AEDGE sensitivity curves at
low frequencies do not take into account the possible effects of instrumental noise.}
We also show for comparison the LISA sensitivity curve, which
is also impacted by the galactic binary background, as well as the sensitivity curves for LIGO and ET.

As stressed in~\cite{AION,AEDGE:2019nxb}, AION-km and AEDGE are ideal for measuring the final merger and ringdown
stages of intermediate-mass black hole binaries at redshifts $\lesssim 10$. These will map out the
merger histories of supermassive black holes, casting light on their origins. Fig.~\ref{fig:strain} 
also illustrates the potential synergies between AION-km/AEDGE measurements and those by LIGO/ET at
higher frequencies and LISA at lower frequencies. Times before mergers at early stages of inspiral
are indicated. We see that AION-km may be able to observe upcoming LIGO-class mergers of pairs of
30 solar mass black holes months beforehand, and that LISA may be able to measure intermediate-mass
black hole mergers years in advance. Combining measurements by different detectors will enable stringent
tests of general relativity, such as probes of the possible mass of the graviton, as discussed in~\cite{Ellis:2020lxl},
for example. Current LIGO/Virgo data provide an upper limit on the gravitino mass of $1.7 \times 10^{-23}$~eV,
observations of a similar event by AION could probe a mass of $1.1 \times 10^{-24}$~eV, which could be 
improved to $1.3 \times 10^{-25}$~eV by AEDGE, and AEDGE observations of a merger of more massive black
holes could improve the sensitivity to $< 10^{-26}$~eV.

\begin{figure}[htb!]
\centering 
\includegraphics[width=6cm]{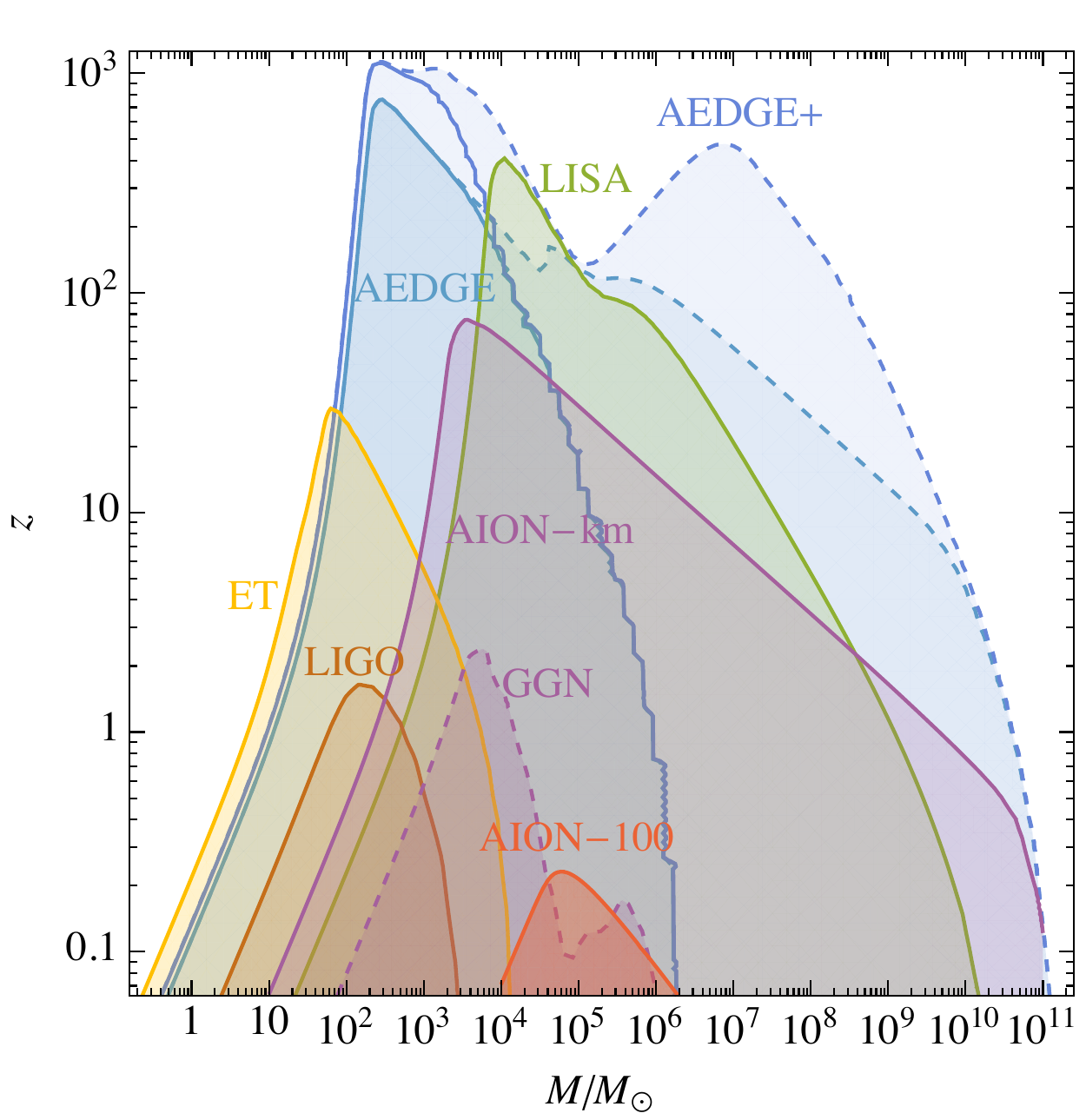} \hspace{2mm}
\includegraphics[width=7cm]{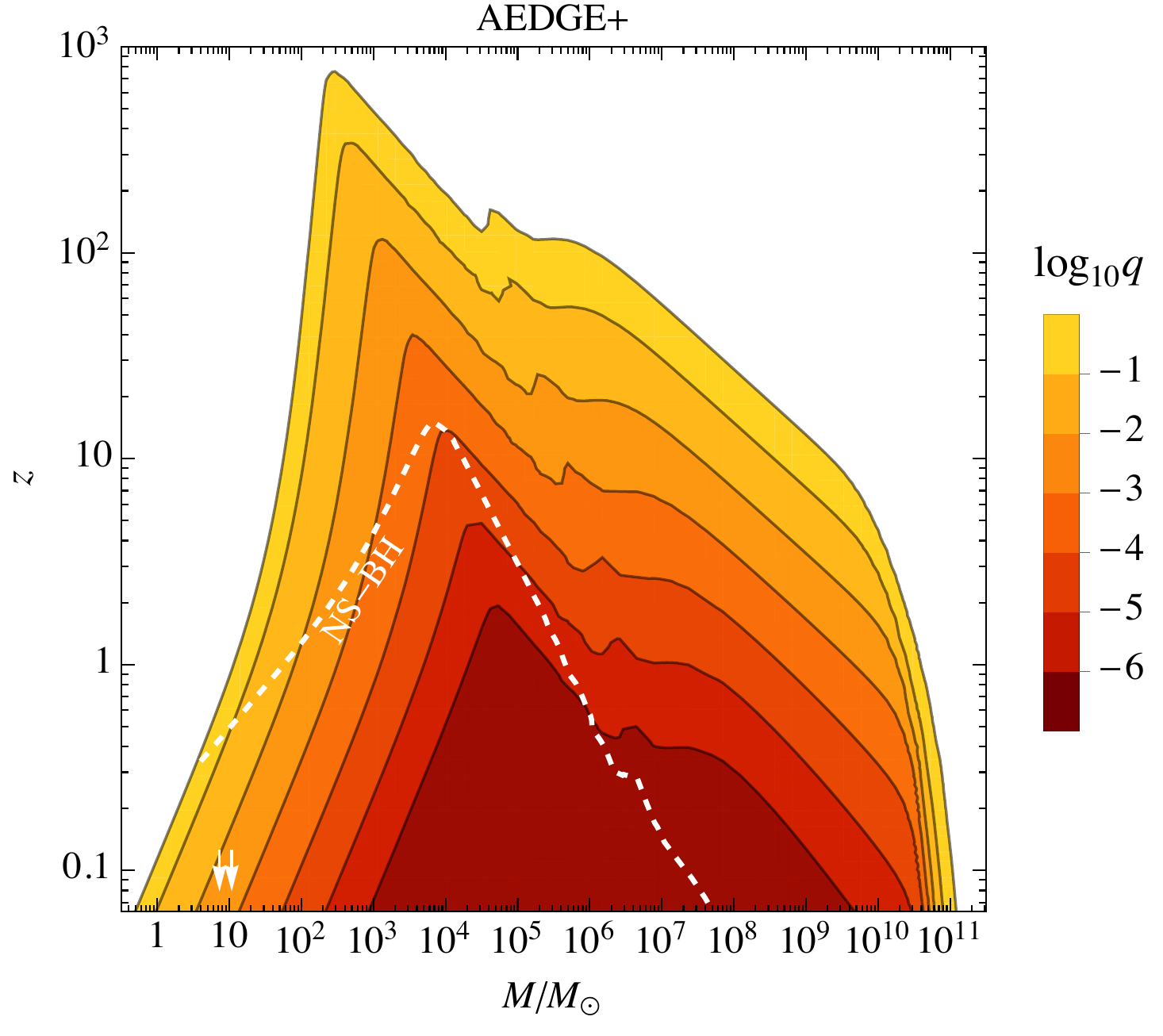}
\caption{\it Left panel: Signal-to-noise ratio (SNR) $= 8$ sensitivities of LIGO, ET, LISA, AION
and AEDGE to equal-mass black hole binaries as functions of the binary total mass and the redshift $z$. Right panel: The SNR $= 8$ sensitivity curves for observations by AEDGE
of unequal-mass mergers, as functions of the heavier black hole mass and the redshift $z$ for
different values of the mass ratio $q$. Black hole/Neutron star mergers correspond to the white
dashed line. The vertical arrows point to the $(M/M_\odot, z)$ values of the two events of this type observed recently~\cite{LIGOScientific:2021qlt}.
}
\label{fig:Mzq}
\end{figure}

The left panel of Fig.~\ref{fig:Mzq} displays the sensitivities of AION, AEDGE, LIGO, ET and LISA 
at the signal-to-noise ratio SNR $= 8$ level to equal-mass binaries as functions of the black hole 
masses and the redshift $z$. AION-km and AEDGE can observe the early stages of mergers observed subsequently with LIGO or ET, whereas LISA could observe early infall stages of mergers
subsequently observed with AION-km or AEDGE. This reemphasises the complementarity of atom
and laser interferometer detectors, and the advantages of operating them together in a network.

The right panel of Fig.~\ref{fig:Mzq} displays the SNR $= 8$ sensitivity of AEDGE with an external atom cloud
to mergers of unequal mass binaries. We see that mergers could be observed at $z > 10$ over a very large range of
masses, extending to $\gtrsim 10^{9}$ solar masses if the instrumental noise
can be mitigated, and including any mergers involving black holes in the expected mass gap around 100 solar masses~\cite{massgap}.
The sensitivity extends to $z > 10^2$ for a range of masses of particular interest for the formation of supermassive
black holes. It is unclear how astrophysical black holes could have formed so early, so such observations
might be a unique window on possible primordial black holes~\cite{Hutsi:2020sol}. In addition, AEDGE is sensitive to very light $<1M_\odot$ binaries, providing a powerful probe of possible sub-solar mass primordial black holes~\cite{Pujolas:2021yaw}.
The sensitivity of AEDGE to black hole/neutron star mergers is illustrated by the dashed line. The two such mergers reported recently by the LIGO, Virgo
and KAGRA Collaborations~\cite{LIGOScientific:2021qlt} would be well within reach of AEDGE, which would be able to extend searches
for such events to much larger redshifts and black hole masses.

\subsection{Supernovae}
\label{sec:SN}

Three-dimensional simulations of stellar collapses to form supernovae (SNe) find turbulence and anisotropies, results that are supported by the observed kicks to neutron stars that are presumably due to
anisotropic emissions of neutrinos.
These emissions can produce GW signals at different stages of SN events. 
One possible source is anisotropic emission in the initial
collapse stage, which would appear at frequencies ${\cal O}(10^3)$~Hz, and hence is of interest to LIGO/Virgo/KAGRA
but out of the AION/AEDGE range. The other is the gravitational memory effect due to asymmetric neutrino emission during the
accretion phase following the collapse~\cite{Epstein:1978dv}, which is expected to peak in a frequency range $\sim 1$~Hz, and hence is of interest to
AION and AEDGE.~\footnote{We do not discuss the possibilities for 
GW emissions in the subsequent cooling phase, which have not yet
been studied numerically. See also~\cite{LP} for a suggestion to search for GW emission from ultra-relativistic jets
that could be generated by some core-collapse SNe.} This effect has recently been studied numerically in~\cite{Mukhopadhyay:2021zbt}, using phenomenological models
inspired by numerical simulations. These models reproduce typical features of the neutrino luminosity and of the 
anisotropy in the neutrino emission over time scales $\gtrsim 0.1$~s, during the accretion phase that follows the initial
neutronisation phase. One well-motivated model (Ac3G) approximates with three Gaussian functions the results of numerical simulations
of a SN collapsing to a neutron star, and the other model (LAc3G) assumes a longer accretion phase, as could be appropriate
for the collapse of a higher-mass star to a black hole.

\begin{figure}
\centering 
\includegraphics[width=6.5cm]{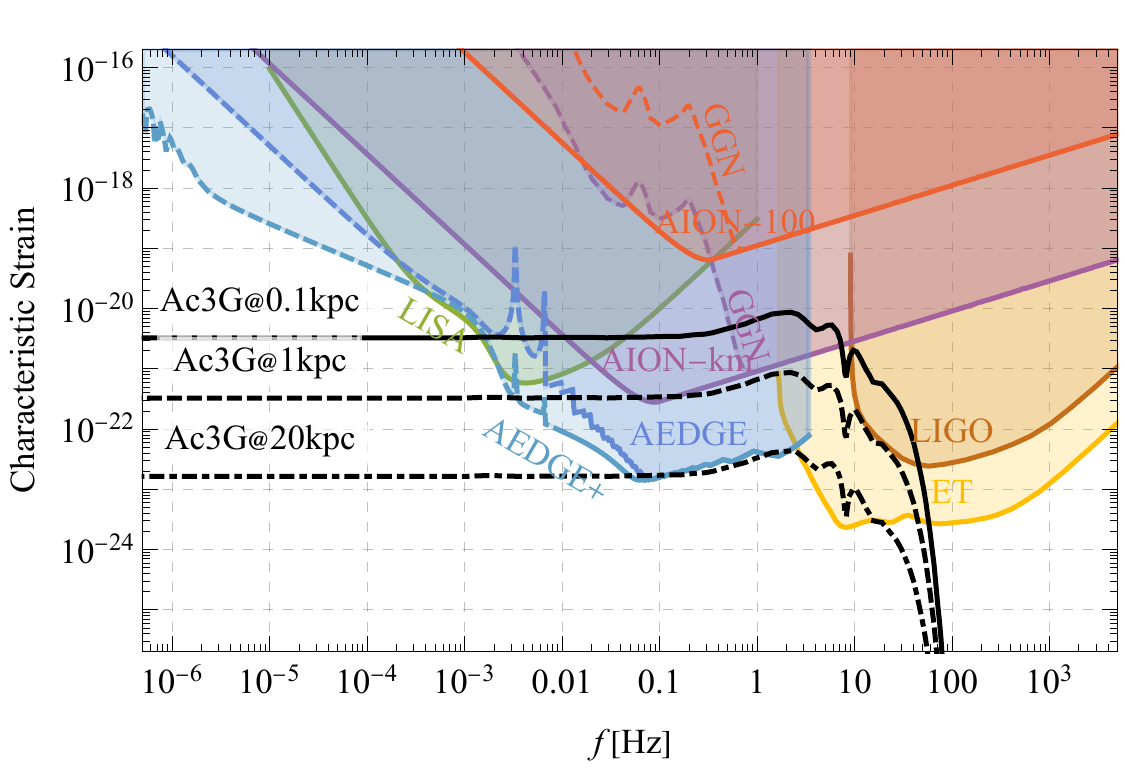}
\includegraphics[width=6.5cm]{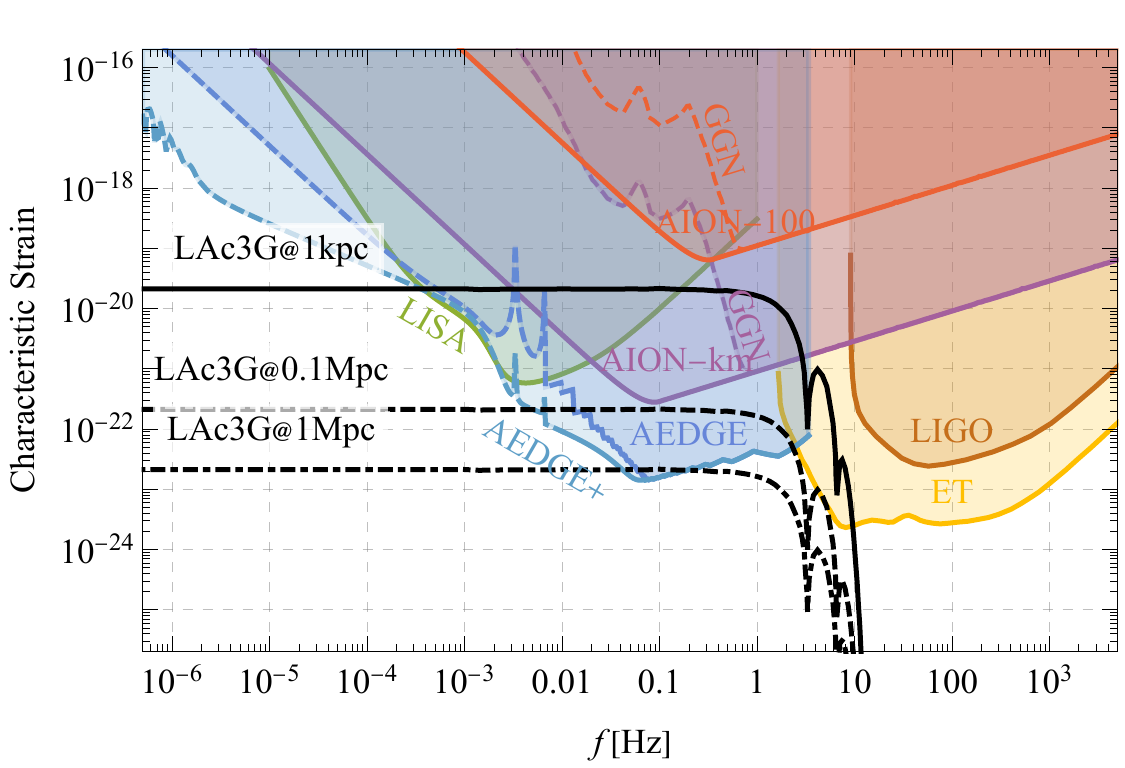}
\caption{\it The strain sensitivities of AION-100 and -km and the two AEDGE variants compared with the neutrino
gravitational memory signal expected from a core-collapse supernova calculated using models Ac3G (left panel) and
LAc3G (right panel) from~\cite{Mukhopadhyay:2021zbt}.}
\label{fig:SN}
\end{figure}

The left panel of Fig.~\ref{fig:SN} compares the AION and AEDGE sensitivity curves with the GW spectrum predicted by the
Ac3G model for the neutrino gravitational memory signal from a SN collapse at a distance of 100~pc (similar to the distances
of nearby Tucana-Horologium~\cite{Mama} and Scorpio-Centaurus~\cite{Breit} stellar associations that may have hosted 
SN explosions within the past few million years~\cite{nearbySN}), 1~kpc and 10~kpc
(a range including the galactic centre). We see that a SN at 100~pc should be detectable by AION-km, whereas a SN at 1~kpc
would be at the limit of AION detectability. However, AEDGE should be able to detect GWs from such a SN out to a distance of $\sim 10$~kpc.
The right panel of Fig.~\ref{fig:SN} shows a similar comparison of GW calculations with the LAc3G model  for events at distances
of 1~kpc, 100~kpc (a range including all of our galaxy and the Magellanic clouds) and 1~Mpc (a range including the Andromeda (M31)
and Triangulum (M33) galaxies). We see that AION might be able to detect GWs from an LAc3G-like event anywhere in our galaxy or the
Magellanic clouds, whereas AEDGE might be able to detect such an event in either M31 or M33. 

\section{Fundamental Physics Sources of GWs}

\subsection{Phase Transitions}
\label{sec:PT}

One of the interesting possibilities for GW studies is the existence
of a stochastic GW background generated by a first-order phase transition
during the early history of the Universe~\cite{Witten:1984rs}, via collisions between bubbles of
the new vacuum, sound waves and turbulence in the plasma~\cite{Caprini:2015zlo,Caprini:2019egz}. 
Neither the electroweak nor the quark-hadron phase transition is expected 
to have been first-order, so any such GW signal would be clear evidence for
physics beyond the Standard Model. The first searches
for such signals in the LIGO/Virgo data were performed
in Ref.~\cite{Romero:2021kby}, and in future more sensitive experiments will probe large fractions of the relevant parameter space.

\begin{figure}
\centering 
\includegraphics[width=0.32\textwidth]{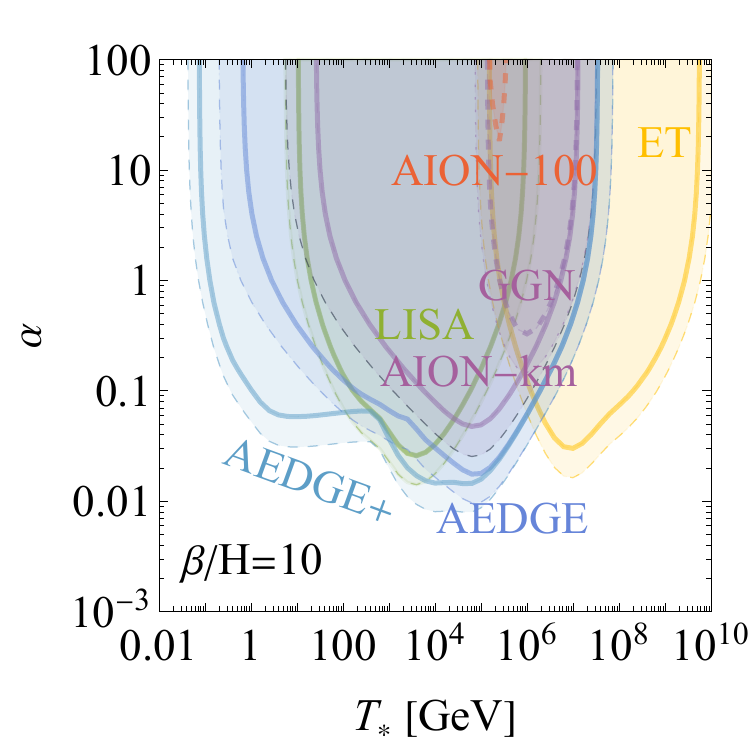}
\includegraphics[width=0.32\textwidth]{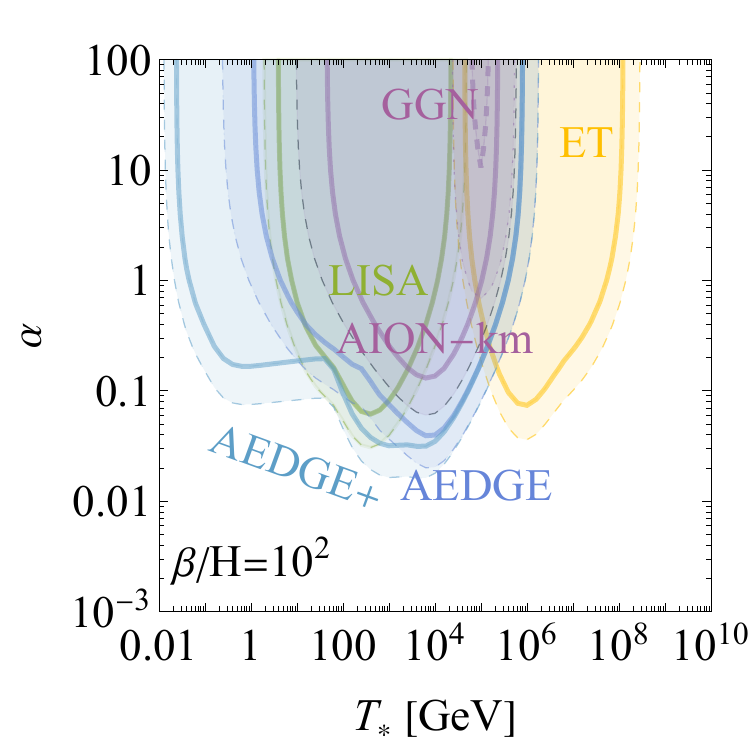}
\includegraphics[width=0.32\textwidth]{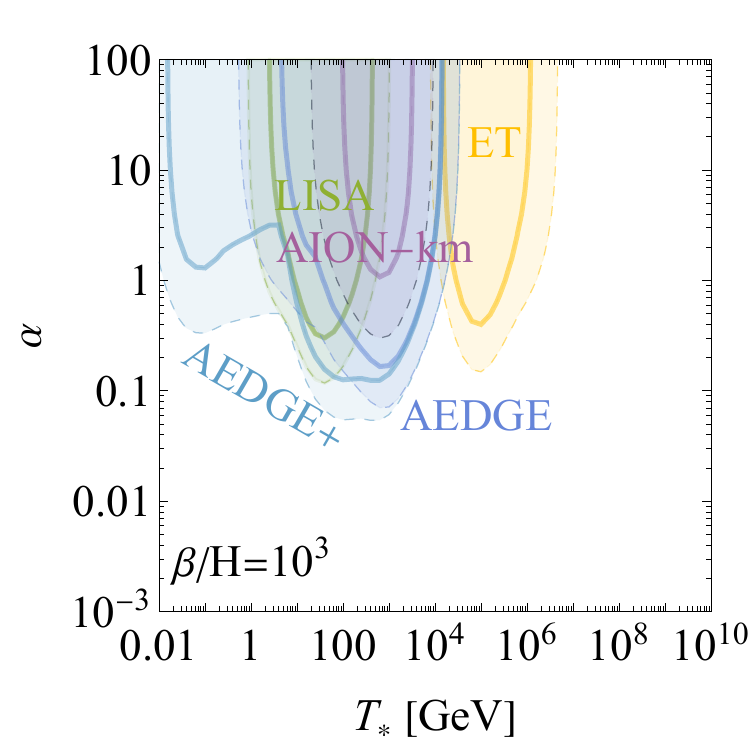}
\caption{\it Sensitivities of AION-100 and -km, AEDGE and AEDGE+ and LISA in the $(T_*, \alpha)$ plane
to GWs from generic first-order transitions with the indicated values of the transition rate $\beta/H$. Thick solid lines show $SNR=10$ while dashed lines $SNR=1$ except for AION-km GGN for which thick dashed line shows $SNR=10$ while dotted line $SNR=1$}
\label{fig:PT}
\end{figure}

It is possible to characterise first-order phase transitions in terms of
two dimensionless parameters, the strength of the phase transition, $\alpha$, and 
the inverse duration of the phase transition, $\beta$, in addition to the
reheating temperature $T_*$. Fig.~\ref{fig:PT} shows the reaches of AION, AEDGE,
LISA and ET for the three representative values $\beta/H = 10$,
$10^2$ and $10^3$ assuming the signal is dominantly produced by sound waves~\cite{Hindmarsh:2015qta,Hindmarsh:2016lnk,Hindmarsh:2017gnf,Caprini:2019egz}.
It is, however, important to remember that due to the fact that the sound wave period only lasts a small fraction of the Hubble time~\cite{Ellis:2018mja,Ellis:2019oqb,Ellis:2020awk} the dependence on key parameters is the same as it would be for supercooled transitions in which bubble collisions would be the dominant signal~\cite{Ellis:2020nnr,Lewicki:2020jiv,Lewicki:2020azd} and as a result the appropriate plot for that source would be very similar. Temperature of the transition effectively controls the frequency of the signal through redshift. Thus as expected 
AION-km has very good sensitivity for temperatures between LISA and ET provided $\alpha \gtrsim {\cal O}(0.1)$ provided $\beta/H\lesssim 10^2$. The $\alpha$ ranges covered by LISA and the in-satellite
AEDGE configuration are quite similar, namely $\alpha = {\cal O}(10^{-2})$, 
while the external cloud version has an extended coverage in temperature (or frequencies).

\begin{figure}
\centering 
\includegraphics[width=6.9cm]{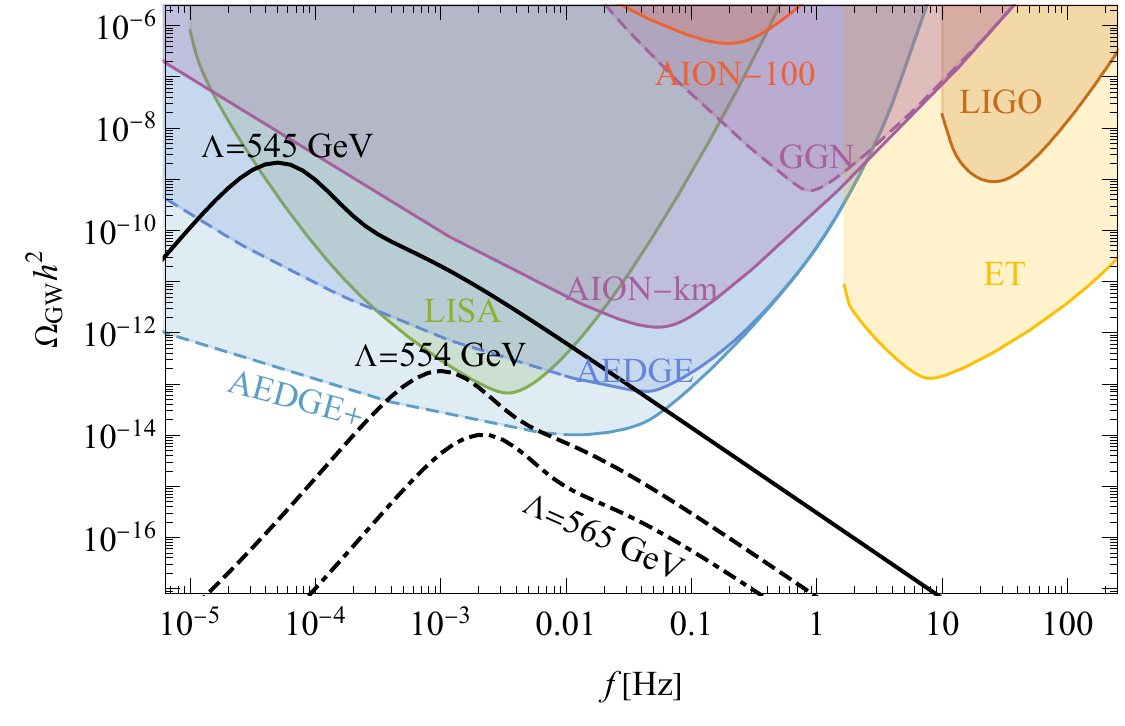}
\includegraphics[width=6.1cm]{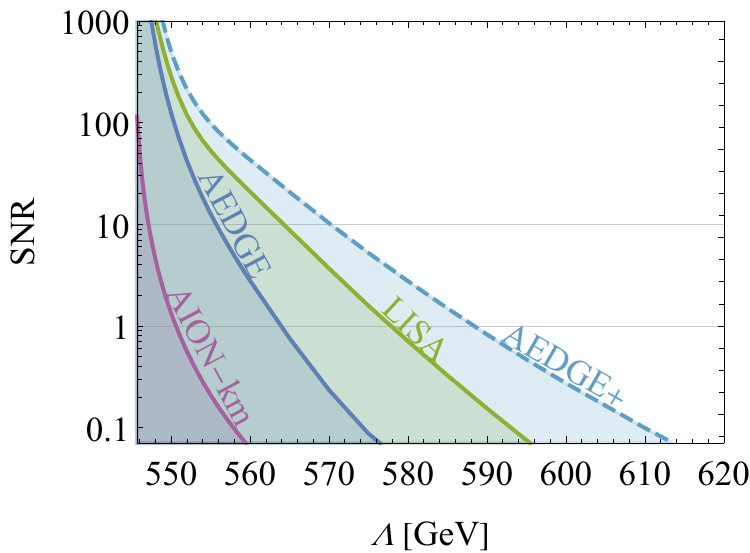} \\
\caption{\it Sensitivities to $\Lambda$ in an extension of the Standard Model including a $H^6/\Lambda^2$
interaction. Left: GW signals for representative values of $\Lambda$ compared with the sensitivities
of AION-100 and -km, LIGO, ET  and LISA. Right: SNR values as functions of $\Lambda$ for LISA, AEDGE
and AEDGE+. 
}
\label{fig:H6}
\end{figure}

It is interesting to look at the GW spectra due to
first-order phase transitions in specific extensions of the Standard Model.
A one-parameter example is illustrated in Fig.~\ref{fig:H6}, namely the Standard Model
supplemented by a dimension-6 $H^6$ operator as might be generated by high-mass
dynamics, scaled by a characteristic mass parameter $\Lambda$~\cite{Grojean:2004xa,Chala:2018ari,Ellis:2018mja}. We see in the left
panel how the GW signal weakens as $\Lambda$ increases, and in the right panel
we see that the sensitivity of LISA reaches up to $\Lambda\sim 565$~GeV for SNR = 10 and surpasses that of in-satellite AEDGE in this particular model due to its relatively low transition temperatures and peak frequencies, reaching $\Lambda\sim 555$~GeV. However, AEDGE+ has considerably
greater sensitivity, to $\sim 575$~GeV for SNR = 10.
Here we also included the contribution to the spectrum produced by turbulence~\cite{Caprini:2009yp,Niksa:2018ofa,Pol:2019yex}, assuming that all of the kinetic energy of the plasma is converted into turbulence at the end of the sound wave period~\cite{Ellis:2019oqb}. As a result, the slower-falling turbulence source takes over at higher frequencies. While this assumption is likely to be too optimistic, the correct value is not currently known~\cite{Caprini:2019egz}, which means that broad coverage of the observed spectrum could allow us to probe microphysical details of the phase transition.
The EFT approach loses accuracy at very low values
of the cut-off scale $\Lambda$. However, we treat it as a toy model to reproduce qualitatively the possible PT dynamics in simple extensions of SM such as that with a (sufficiently heavy) singlet scalar~\cite{Postma:2020toi}. This is the spirit in which the comparisons between different experimental sensitivities in Fig.~\ref{fig:H6} should be interpreted.

\subsection{Cosmic Strings}
\label{sec:strings}

\begin{figure}
\centering 
\includegraphics[height=4.1cm]{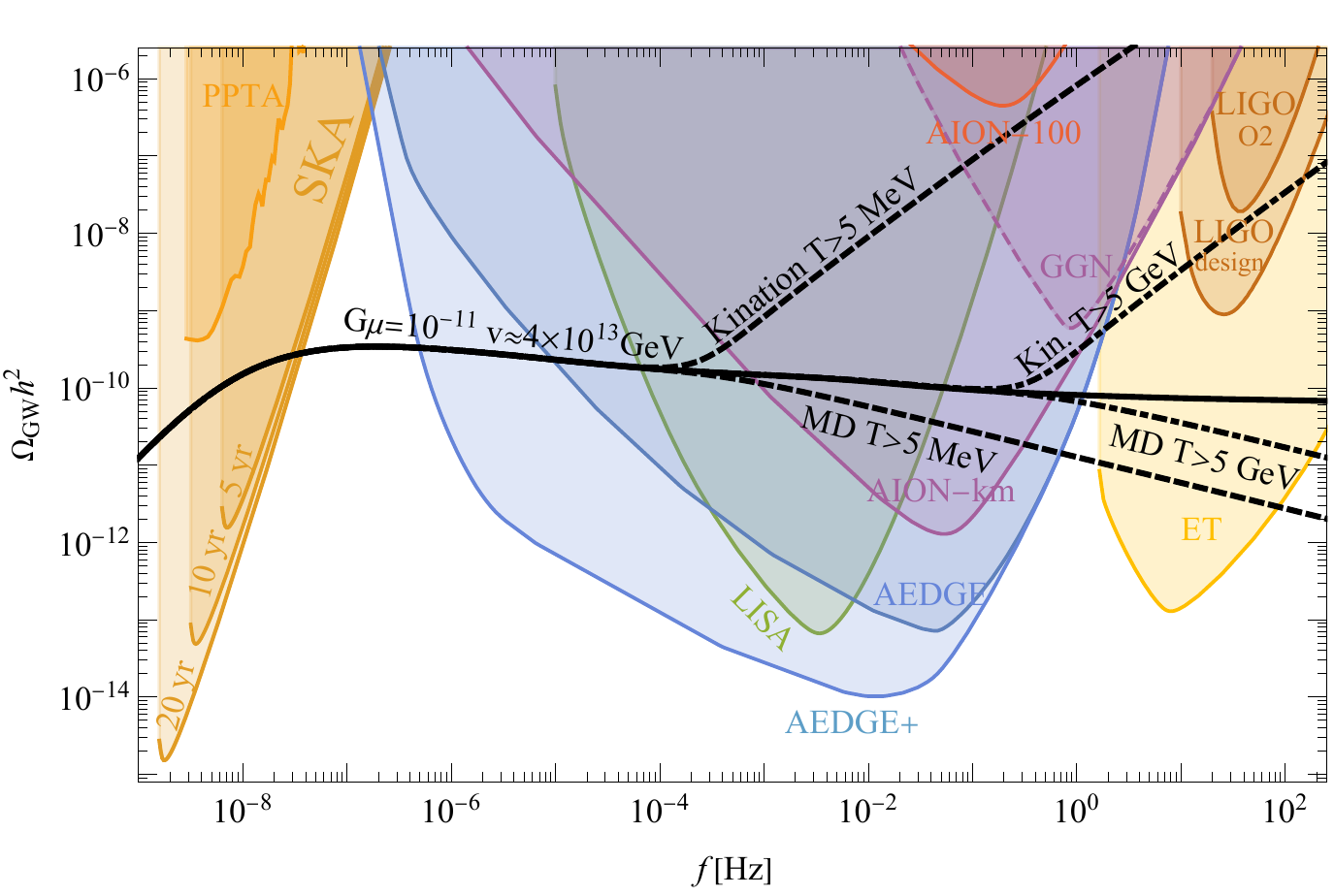}
\includegraphics[height=4.3cm]{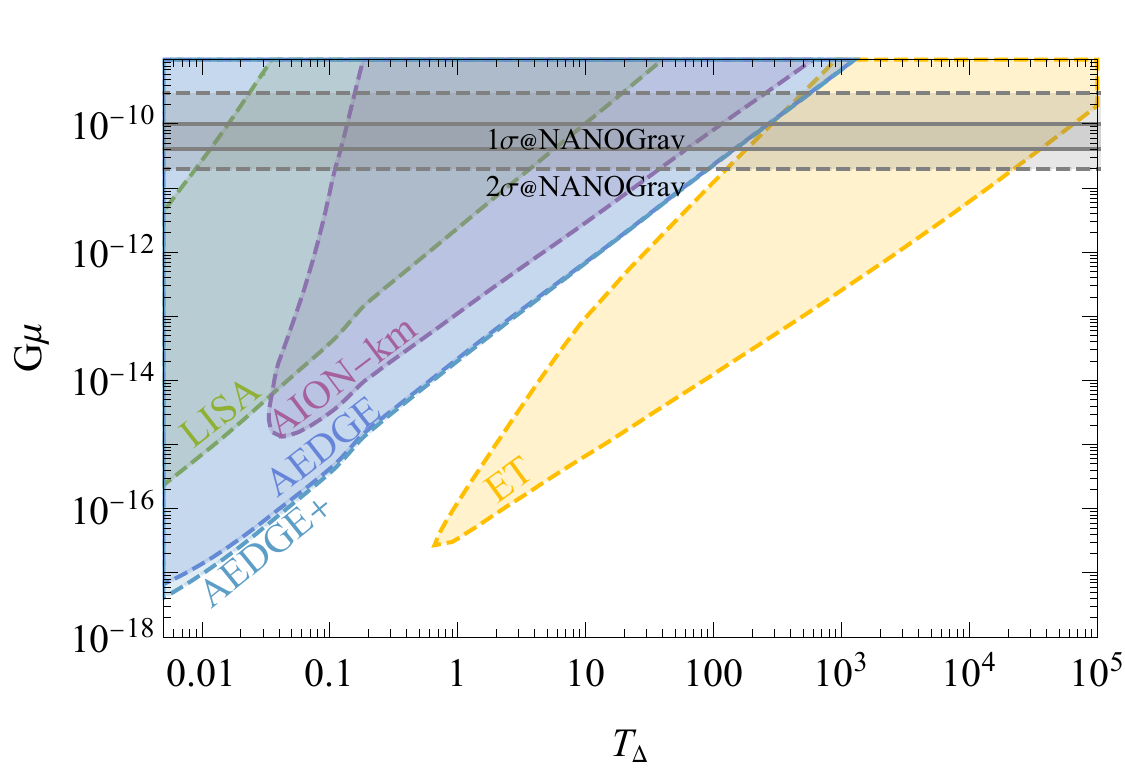}
\caption{\it Sensitivities of cosmic string measurements to modifications of the cosmological expansion rate.
Left panel: Kination or
matter dominance (MD) at temperatures $T > 5$~MeV or $5$~GeV.
 Right panel: Sensitivity to modification of the expansion rate at a temperature $T_\Delta$ for a given value of the string tension $G\mu$.}
\label{fig:expansion}
\end{figure}

The GW radiation from loops produced by a cosmic string network could provide a stochastic GW background signal observable over 
many decades in frequency~\cite{Blanco-Pillado:2017oxo,Blanco-Pillado:2017rnf,Auclair:2019wcv}. Measurements by pulsar timing arrays (PTAs) at frequencies $\lesssim 10^{-8}$~Hz currently provide 
the strongest upper limit on the cosmic string tension $G_\mu$ of around $10^{-10}$, and a tantalising measurement by
the NANOGrav PTA of a stochastic common-spectrum process~\cite{NANOGrav:2020bcs} could be evidence for GWs, which might come from cosmic strings with
$G\mu \sim 4 \times 10^{-10} - 10^{-11}$~\cite{Ellis:2020ena,Blasi:2020mfx,Blanco-Pillado:2021ygr}.~\footnote{Results
from the Parkes PTA that are consistent with those of NANOGrav have recently been
reported~\cite{Goncharov:2021oub}, but their interpretation remains open.}
The cosmic string interpretation would suggest that the signal should also be measurable in LISA, AION-km, AEDGE
and ET, though not in LIGO/Virgo/KAGRA~\cite{Ellis:2020ena}. Comparisons of the magnitudes of the GW signals from cosmic strings in different
frequency bands could provide unique insights into the expansion history of the early Universe~\cite{Cui:2017ufi,Cui:2018rwi,Cui:2019kkd,Gouttenoire:2019kij,Gouttenoire:2019rtn}, as illustrated in
Fig.~\ref{fig:expansion}.
The left panel of Fig.~\ref{fig:expansion} shows that AION-km and
AEDGE measurements could detect modification of the expansion rate at temperatures
$T > 5$~GeV, and hence be able to cast light on any possible discrepancy between the strengths of the GW signals in ET and LISA. For illustration we chose rather drastic modifications consisting of an epoch of matter domination (MD) or kination (Kin.), but even much more subtle effects coming, e.g., from the appearance of additional relativistic degrees of freedom in a dark sector would also be observable.
The growing signal boosted by kination must be cut off at some point (within the Figure for the 5~MeV line, but possibly beyond it for the 5~GeV line), so as to obey the BBN bounds on additional energy stored in GWs~\cite{Smith:2006nka}. The cut-off depends on what preceded the kination epoch, and we do not discuss here the model-building issues that arise. However, one should be aware this effect imposes a limit on the possible duration of the kination period.
The right panel shows the sensitivities of various GW experiments to modification of the expansion rate at a temperature $T_\Delta$ thanks to feature appearing in the spectrum produced by a cosmic string network with string tension $G\mu$. We see that, for $G\mu$ in the range corresponding to the recent NANOGrav data, AION-km and AEDGE cover an important range of $T_\Delta \sim 10 - 100$~GeV that is not accessible to LISA and ET. 

\begin{figure}
\centering 
\includegraphics[width=9.65cm]{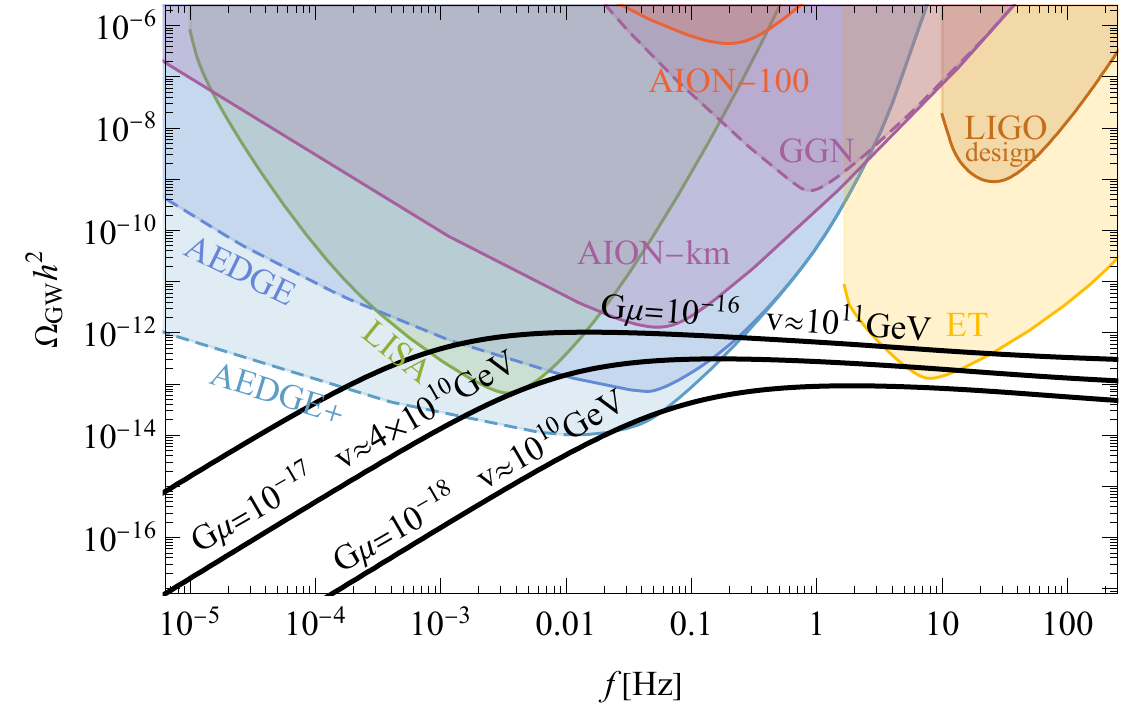}
\vspace{-3mm}
\caption{\it Sensitivities to the cosmic strings with tension $G \mu$ of AION-100 and -km, 
AEDGE and AEDGE+, LIGO, ET and LISA.}
\label{fig:Strings}
\end{figure}

The above examples illustrate what AION and AEDGE could contribute if the NANOGrav indication of a possible
GW signal is confirmed and due to cosmic strings. However, this interpretation of the NANOGrav data may
not be confirmed, and the cosmic string tension $G \mu$ may lie below the range discussed in the previous paragraph.
As we see in Fig.~\ref{fig:Strings}, AION-km and the two versions of AEDGE would be sensitive to much smaller values of $G \mu$, namely
$G\mu \sim 10^{-16} (10^{-17}) (10^{-18})$ for AION-km, the standard version of AEDGE and AEDGE+, respectively.

\subsection{Inflationary Fluctuations and Primordial Black Holes}
\label{sec:fluctuations}

The favoured scenario for primordial black hole (PBH) formation relies on high-amplitude quantum curvature fluctuations 
generated during cosmological inflation~\cite{Carr:1974nx}. Such fluctuations would also source GWs at second order,
generating a scalar induced GW 
background~\cite{Matarrese:1993zf,Matarrese:1997ay,Nakamura:2004rm,Ananda:2006af,Baumann:2007zm}. The first searches for such a GW background in the LIGO/Virgo data have already been performed~\cite{Kapadia:2020pnr,Romero-Rodriguez:2021aws}, and in the future more sensitive detectors will provide powerful probes of the favoured PBH formation mechanism. 

\begin{figure}
\centering 
\includegraphics[width=0.65\textwidth]{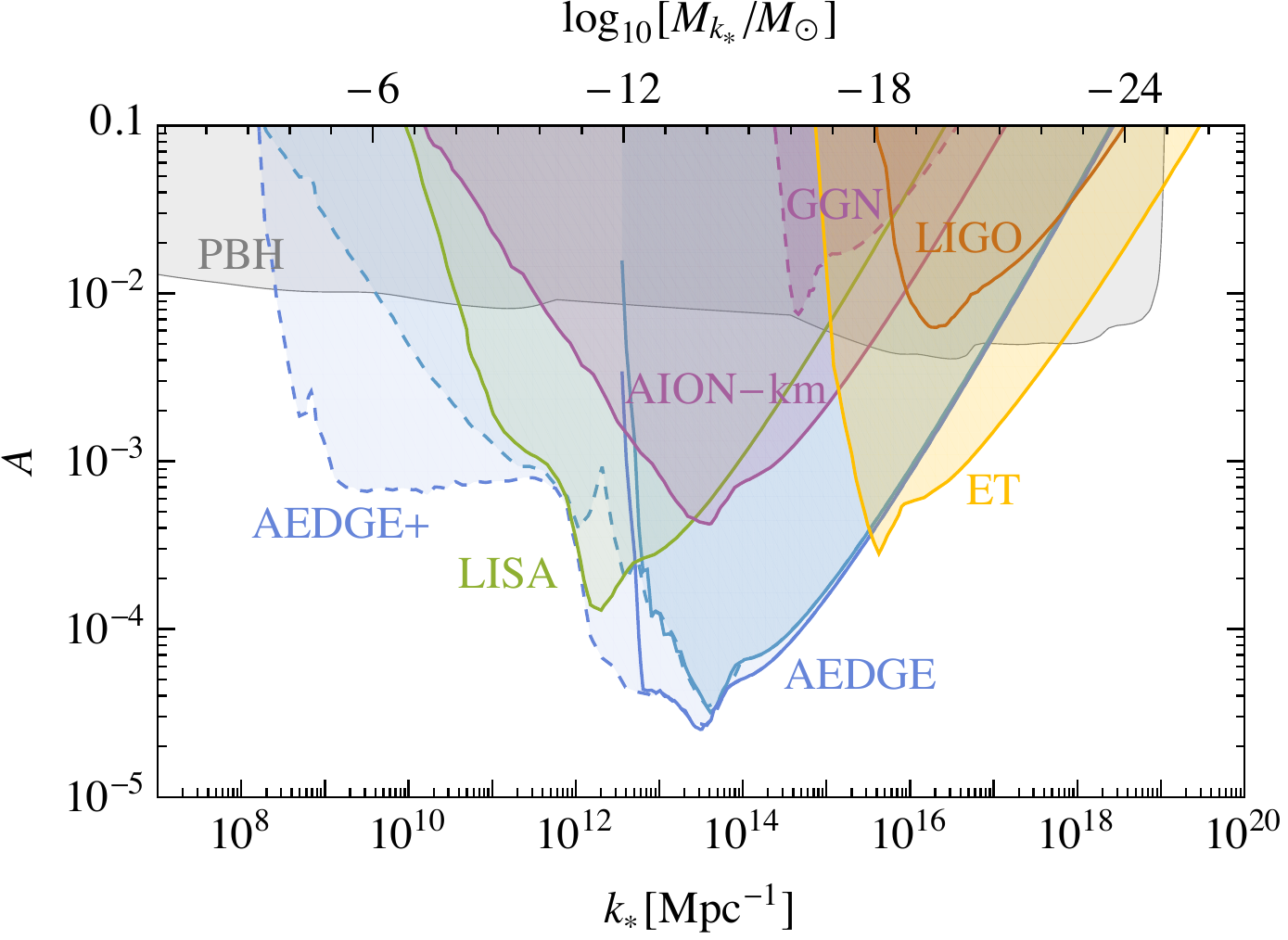}
\vspace{-3mm}
\caption{\it Sensitivities to the amplitude $A$ of a delta function peak in the curvature power spectrum at scale $k_*$ of AION-km, 
AEDGE and AEDGE+, LIGO, ET and LISA. The gray region is excluded by constrains on primordial black holes. The top axis indicates the horizon mass that corresponds 
approximately to the mass of the formed primordial black holes.}
\label{fig:PBHs}
\end{figure}

In the 
standard radiation-dominated early Universe, the amplitude of the curvature fluctuations would need to be 
$A \sim 0.01$ for the formation of a significant PBH abundance. As shown in Fig.~\ref{fig:PBHs},
such fluctuations would generate a GW background that is well within the reach of future GW observatories. 
Assuming that the GGN can be mitigated, AION-km could probe this mechanism for PBH formation in the range 
from $\sim 10^{-10}M_\odot$ to $\sim 10^{-18}M_\odot$ and AEDGE could reach down to $\sim 10^{-21}M_\odot$. 
In particular, both of these experiments can probe the formation of PBHs with masses close to those of asteroids,
where the current PBH constraints~\cite{Carr:2020xqk} would allow PBHs to constitute all the dark matter. 
The sensitivities of AION and AEDGE extend down to fluctuation amplitudes $A\sim 10^{-3} - 10^{-4}$, 
which are too small to form a significant density of PBHs, but have broader relevance for studies of 
inflation models, providing, for example, powerful probes of thermal inflation~\cite{Lewicki:2021xku}.

\section{Ultralight Dark Matter}

\begin{figure}
\centering 
\includegraphics[width=6.5cm]{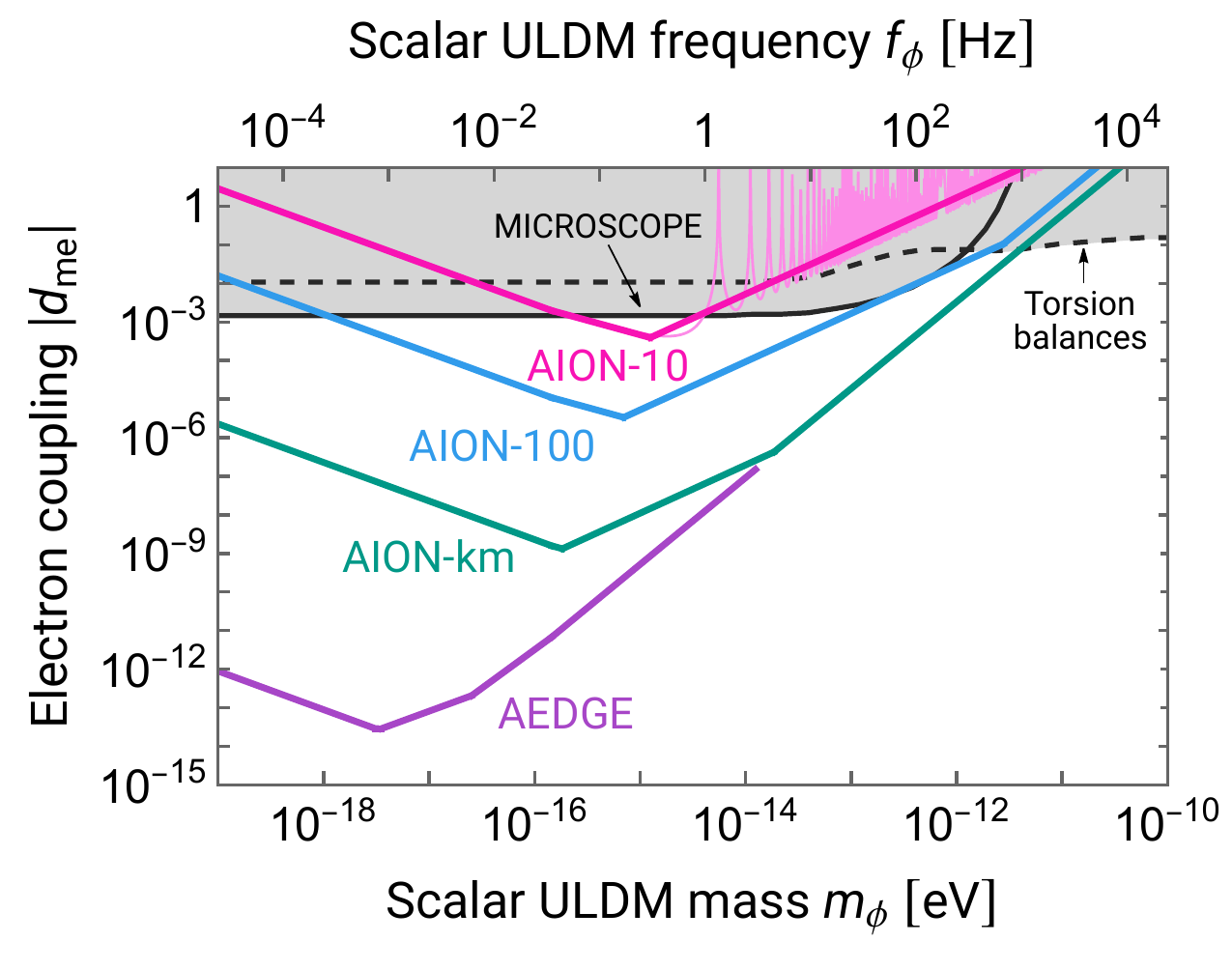}
\includegraphics[width=6.5cm]{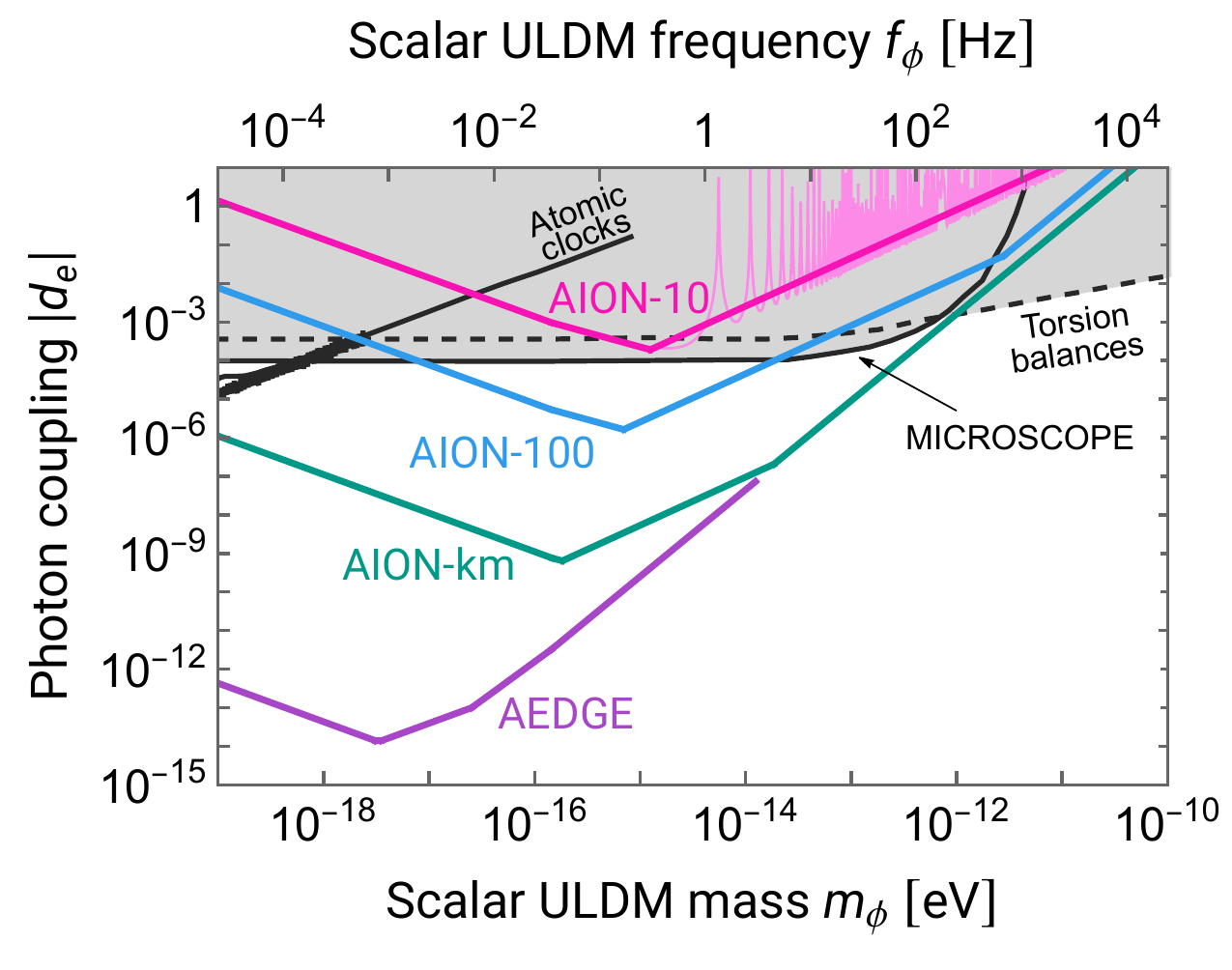}
\vspace{-3mm}
\caption{\it Sensitivity projections to ULSDM linearly coupled to electrons (left panel) and photons (right panel). The lighter-pink AION-10 curve shows the oscillatory nature of the sensitivity projections, while the darker-pink straight AION-10 curve shows the envelope of the oscillations. For clarity, for AION-100, AION-km and AEDGE, we only show the envelope. The shaded grey region shows the existing constraints from searches for violations of the equivalence principle with torsion balances~\cite{Wagner:2012ui}, atomic spectroscopy~\cite{Hees:2016gop} and the MICROSCOPE experiment~\cite{Berge:2017ovy}.
}
\label{fig:AIONULDM}
\end{figure}

The primary focus of this paper has been on the GW science topics that can be addressed with the AION and AEDGE projects.
However, the same AI instruments also have unique capabilities to detect ultralight dark matter candidates, and 
proposals have been made for probing scalar, pseudoscalar and vector fields of dark matter~\cite{Graham:2015ifn, Arvanitaki:2016fyj}.
In this final Section, we provide a brief review of the capabilities of the various stages of AION and AEDGE for detecting ultralight scalar dark matter (ULSDM) that is linearly coupled to Standard Model fields.

The linear couplings of ULSDM to the electron and photon fields induce an effective time-dependent correction to the electron mass and electromagnetic fine-structure constant, respectively~\cite{Stadnik:2014tta}. In turn, the time variations of these parameters lead to a small time-dependence in the energy-level of atoms, including in the `clock' transition of strontium, which is the isotope proposed for AION and AEDGE.
The amplitude of the ULSDM-induced transition-energy oscillation depends on the ULSDM energy density at the position of the detector, the ULSDM mass, and the strength of the linear coupling between the ULSDM and electron or photon fields. The frequency of the transition-energy oscillation is set by the ULSDM mass and the ULSDM-induced signal is largest when this frequency lies in the mid-frequency band 
in which AION and AEDGE will operate.

The left (right) panel of Fig.~\ref{fig:AIONULDM} shows sensitivity projections for ULSDM linearly coupled to electrons (photons) for a SNR = 1, using the procedure outlined in Ref.~\cite{BadurinaRefinedDM}. The AION and AEDGE sensitivity curves are compared to existing constraints, shown by the shaded grey regions. The AI sensitivity oscillates as a function of the ULSDM mass, as shown by the light-pink AION-10 curve in both panels of Fig.~\ref{fig:AIONULDM}. However, for clarity, it is often only the envelope of the oscillations that is plotted, and we have followed this procedure for the AION-100, AION-km and AEDGE projections. Also, for clarity we have plotted the AEDGE sensitivity curve only to the point where it approaches the AION-km line even though the sensitivity extends to higher frequencies: if plotted, the extension of the AEDGE sensitivity curve would lie on top of the AION-km line.

Figure~\ref{fig:AIONULDM} shows the exciting prospects for AI detectors of exploring unconstrained domains of parameter space in both the couplings to the Standard Model fields, and for ULSDM masses between $10^{-18}$~eV and $10^{-12}$~eV. AION-10 hopes to approach or even surpass existing constraints, while AION-100 and AION-km should significantly extend the reach to lower values of the coupling. These projections assume that the phase-noise is limited by atom shot-noise. Below around 0.1~Hz, it is expected that GGN will start to dominate, and we have not extrapolated the sensitivities of the terrestrial experiments below this frequency. As space-borne experiments do not have to contend with the same GGN noise, we have extended the AEDGE projections to lower frequencies, or equivalently, to lower values of the ULSDM mass that are complementary to the parameters that can be tested with terrestrial AIs.

\section{Conclusions}
\label{sec:conx}

Atom interferometry is a promising technique for many other studies in fundamental physics,
including searches for ultralight dark matter, probes of the weak equivalence
principle and tests of quantum mechanics, as well as the searches for gravitational
waves discussed here. This technique is now emerging from the laboratory to be
deployed in large-scale experiments at the 10 to 100m scale. Ideas are being
developed for possible future experiments at the km scale and in space.
In this article we have reviewed the possible scientific capabilities of such atom
interferometer experiments, focusing on the terrestrial AION project~\cite{Badurina:2019hst} 
and its possible evolution towards a space-borne project called AEDGE~\cite{AEDGE:2019nxb}.

We have discussed the possible sensitivities of AION and AEDGE to gravitational waves over a range
of frequencies from ${\cal O}(1)$~Hz down to ${\cal O}(10^{-6})$~Hz, comparing them with
the sensitivities of the operating LIGO, Virgo and KAGRA detectors, the planned LISA
experiment and the proposed ET experiment. AION and AEDGE have high sensitivities in the
mid-frequency band between the maximum sensitivities of LIGO/Virgo/KAGRA/ET and LISA.
They may also have interesting sensitivities at frequencies below the optimal frequency
of LISA, if instrumental noise can be controlled sufficiently.

We have shown that AION and AEDGE have interesting capabilities for measuring gravitational
waves from the mergers of intermediate mass black holes, casting light on the formation of
the supermassive black holes in the centres of galaxies. Their mid-frequency measurements
would complement the measurements of higher- and lower-frequency detectors, and their
combinations could be used to probe deviations from general relativity in the propagation
of gravitational waves, e.g., by constraining the mass of the graviton. AION and AEDGE
could also measure the gravitational memory effect generated by neutrinos emerging
anisotropically from a supernova. Their measurements could also probe gravitational
waves from a phase transition in the early Universe, as could occur in many extensions 
of the Standard Model. Finally, AION and AEDGE would be sensitive to gravitational
waves emitted by cosmic strings, and studies of their spectrum could reveal non-adiabatic
features in the expansion of the early Universe.

The road towards large-scale atom interferometers is strewn with technical hurdles,
but the potential gravitational-wave capabilities that we have outlined motivate a
a sustained effort to overcome them, so as to reveal fascinating and novel aspects of the
Universe. The unique capabilities of such detectors
to search for ultralight dark matter add to their
prospective interest.

\ack{
The work of O.B was supported by the UK STFC Grant ST/T006994/1 and that of J.E.\ and C.M.\ by the UK STFC Grant ST/T00679X/1
in the framework of the AION Consortium.
The work of J.E.\ was also supported by the UK STFC Grant ST/T000759/1 and by the Estonian Research Council grant MOBTT5. C.M.\ is also supported by the UK STFC Grant ST/N004663/1. L.B.\  is a recipient of an STFC studentship.
The work of M.L.\ was supported by the Polish National Science Center grant 2018/31/D/ST2/02048 and by the Polish National Agency for Academic Exchange within the Polish Returns Programme under agreement PPN/PPO/2020/1/00013/U/00001. The work of V.V.\
was supported by the Spanish MINECO grants FPA2017-88915-P and SEV-2016-0588, the grant
2017-SGR-1069 from the Generalitat de Catalunya, the European Regional Development Fund through the CoE program grant TK133, the Mobilitas Pluss grants MOBTP135 and MOBTT5, and the Estonian Research Council grant PRG803. IFAE is partially funded by the CERCA program of the Generalitat de Catalunya.}

\bibliographystyle{JHEP}
\bibliography{refs}

\end{document}